\documentclass[acmtog]{acmart}

\usepackage{pbox}
\usepackage{booktabs} 
\usepackage{soul}
\usepackage{tikz}
\usepackage{float}
\usepackage{calc}
\usepackage{adjustbox}
\usepackage{float}
\usepackage{multirow}
\usepackage[percent]{overpic}
\usepackage{array}
\usepackage{stackengine}
\usetikzlibrary{shadows.blur}
\usetikzlibrary{positioning}
\usetikzlibrary{shapes.symbols}

\setcitestyle{square}

\usepackage[ruled]{algorithm2e} 
\usepackage{amsmath}

\SetAlFnt{\small}
\SetAlCapFnt{\small}
\SetAlCapNameFnt{\small}
\SetAlCapHSkip{0pt}
\IncMargin{-\parindent}

\acmJournal{TOG}
\acmVolume{X}
\acmNumber{X}
\acmArticle{X}
\acmYear{2018}
\acmMonth{1}

\setcopyright{acmcopyright}

\acmDOI{0000001.0000001_2}

\received{September 2017}
\received{February 2018}

\begin{document}
\title{Exposure: A White-Box Photo Post-Processing Framework}

\author{Yuanming Hu}
\orcid{1234-5678-9012-3456}
\affiliation{%
\institution{Microsoft Research \& MIT CSAIL}
}
\author{Hao He}
\orcid{1234-5678-9012-3456}
\affiliation{%
\institution{Microsoft Research \& MIT CSAIL}
}
\author{Chenxi Xu}
\orcid{1234-5678-9012-3456}
\affiliation{%
\institution{Microsoft Research \& Peking University}
}
\author{Baoyuan Wang}
\affiliation{%
\institution{Microsoft Research}
}
\author{Stephen Lin}
\affiliation{%
\institution{Microsoft Research}
}

\renewcommand\shortauthors{Hu, Y. et al}
\newcommand{\pdfbar}[1]{\tikz \fill[red!40] (0,0) rectangle (#1 ex, 1.0 ex);}
\newcommand{\pdfbarSelected}[1]{\tikz \fill[red!90] (0,0) rectangle (#1 ex, 1.0 ex);}

\newcommand{\ifolder}{2791}

\newcommand{\DeltaX}{7.3}
\newcommand{\DeltaY}{9.0}

\begin{teaserfigure}
\centering
  \resizebox{1.0\linewidth}{!}{
    \begin{tikzpicture}[scale=0.96]
        \node[inner sep=0pt] (final) at (14.0, -5.8) {\includegraphics[width=.83\textwidth]{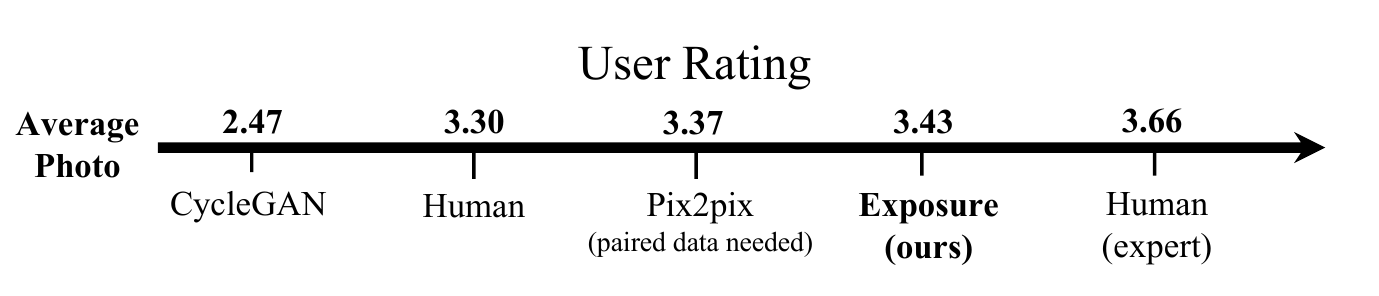}};
        \node[inner sep=0pt] (image1) at (0.7,0) {\includegraphics[width=.19\textwidth]{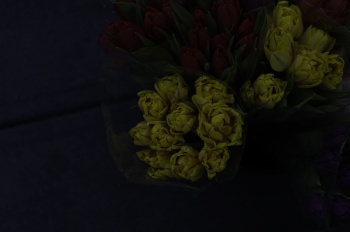}};
        \node[inner sep=0pt] (image2) at (0.7,-3) {\includegraphics[width=.19\textwidth]{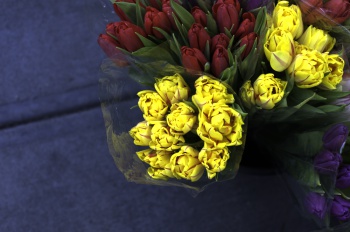}};
        \node[inner sep=0pt] (image3) at (0.7,-6) {\includegraphics[width=.19\textwidth]{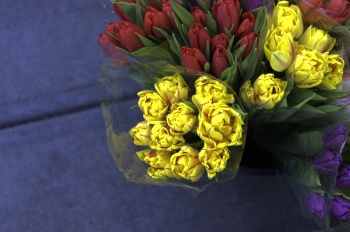}};
        \node[inner sep=0pt] (image4) at (8,0) {\includegraphics[width=.19\textwidth]{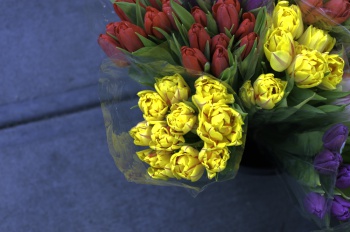}};
        \node[inner sep=0pt] (image5) at (8, -3) {\includegraphics[width=.19\textwidth]{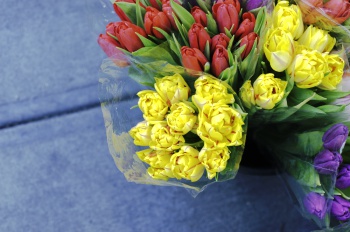}};
        \node[inner sep=0pt] (final) at (17.4, -1.48) {\includegraphics[width=.425\textwidth]{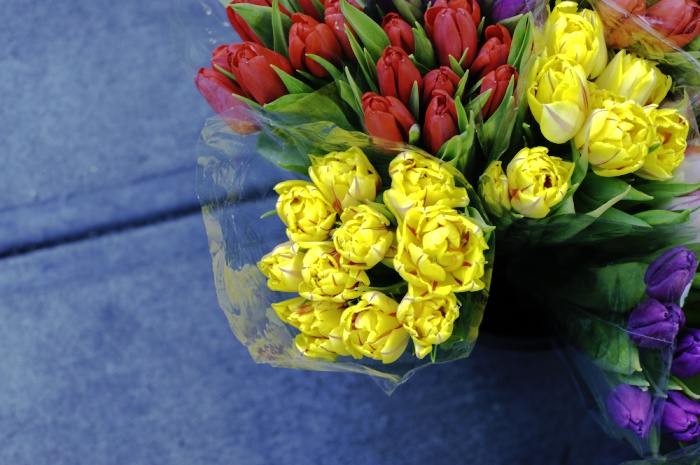}};
\node[draw, rectangle, thick,minimum height=7em,minimum width=7em](agent1) at (4.000000,0.000000) {};
\node (agent1s) at ([yshift=1.4em]agent1.center) {
    \scalebox{0.7}{
    \begin{tabular}{|p{0.5cm}p{0.2cm}p{0.5cm}p{0.2cm}|}
        \hline
        Expo. & \pdfbarSelected{1.148} & Tone & \pdfbar{0.076} \\
        Gam. & \pdfbar{0.129} & Cst. & \pdfbar{0.912} \\
        W.B. & \pdfbar{0.337} & BW & \pdfbar{0.162} \\
        Satu. & \pdfbar{0.054} & Color & \pdfbar{0.180} \\
        \hline
    \end{tabular}
    }
};
\node (agent1d) at ([yshift=-2.0em]agent1.center)
{Exposure $+2.15$};
\node[draw, rectangle, thick,minimum height=7em,minimum width=7em](agent2) at (4.000000,-3.000000) {};
\node (agent2s) at ([yshift=1.4em]agent2.center) {
    \scalebox{0.7}{
    \begin{tabular}{|p{0.5cm}p{0.2cm}p{0.5cm}p{0.2cm}|}
        \hline
        Expo. & \pdfbar{0.019} & Tone & \pdfbar{0.342} \\
        Gam. & \pdfbar{0.133} & Cst. & \pdfbarSelected{0.844} \\
        W.B. & \pdfbar{0.516} & BW & \pdfbar{0.143} \\
        Satu. & \pdfbar{0.517} & Color & \pdfbar{0.486} \\
        \hline
    \end{tabular}
    }
};
\node (agent2d) at ([yshift=-2.0em]agent2.center)
{Contrast $-0.59$};
\node[draw, rectangle, thick,minimum height=7em,minimum width=7em](agent3) at (4.000000,-6.000000) {};
\node (agent3s) at ([yshift=1.4em]agent3.center) {
    \scalebox{0.7}{
    \begin{tabular}{|p{0.5cm}p{0.2cm}p{0.5cm}p{0.2cm}|}
        \hline
        Expo. & \pdfbar{0.021} & Tone & \pdfbar{0.349} \\
        Gam. & \pdfbar{0.070} & Cst. & \pdfbar{0.020} \\
        W.B. & \pdfbar{0.311} & BW & \pdfbar{0.049} \\
        Satu. & \pdfbar{0.135} & Color & \pdfbarSelected{2.045} \\
        \hline
    \end{tabular}
    }
};
\node (agent3d) at ([yshift=-2.0em]agent3.center)
{Color\quad\quad\quad\quad};
\draw[<->] (4.000000,-6.220000) -- (4.000000,-7.100000) -- (4.880000,-7.100000);
\draw[red,-] (4.000000,-7.100000) -- (4.100000,-7.002888);
\draw[red,-] (4.100000,-7.002888) -- (4.200000,-6.894773);
\draw[red,-] (4.200000,-6.894773) -- (4.300000,-6.784917);
\draw[red,-] (4.300000,-6.784917) -- (4.400000,-6.687029);
\draw[red,-] (4.400000,-6.687029) -- (4.500000,-6.593401);
\draw[red,-] (4.500000,-6.593401) -- (4.600000,-6.500069);
\draw[red,-] (4.600000,-6.500069) -- (4.700000,-6.401358);
\draw[red,-] (4.700000,-6.401358) -- (4.800000,-6.300000);\draw[green,-] (4.000000,-7.100000) -- (4.100000,-6.987458);
\draw[green,-] (4.100000,-6.987458) -- (4.200000,-6.876407);
\draw[green,-] (4.200000,-6.876407) -- (4.300000,-6.770038);
\draw[green,-] (4.300000,-6.770038) -- (4.400000,-6.675891);
\draw[green,-] (4.400000,-6.675891) -- (4.500000,-6.583008);
\draw[green,-] (4.500000,-6.583008) -- (4.600000,-6.490012);
\draw[green,-] (4.600000,-6.490012) -- (4.700000,-6.395701);
\draw[green,-] (4.700000,-6.395701) -- (4.800000,-6.300000);\draw[blue,-] (4.000000,-7.100000) -- (4.100000,-6.996136);
\draw[blue,-] (4.100000,-6.996136) -- (4.200000,-6.885276);
\draw[blue,-] (4.200000,-6.885276) -- (4.300000,-6.774281);
\draw[blue,-] (4.300000,-6.774281) -- (4.400000,-6.667178);
\draw[blue,-] (4.400000,-6.667178) -- (4.500000,-6.575259);
\draw[blue,-] (4.500000,-6.575259) -- (4.600000,-6.483448);
\draw[blue,-] (4.600000,-6.483448) -- (4.700000,-6.392174);
\draw[blue,-] (4.700000,-6.392174) -- (4.800000,-6.300000);
\node[draw, rectangle, thick,minimum height=7em,minimum width=7em](agent4) at (\DeltaX+4.000000,\DeltaY-9.000000) {};
\node (agent4s) at ([yshift=1.4em]agent4.center) {
    \scalebox{0.7}{
    \begin{tabular}{|p{0.5cm}p{0.2cm}p{0.5cm}p{0.2cm}|}
        \hline
        Expo. & \pdfbar{0.021} & Tone & \pdfbarSelected{1.597} \\
        Gam. & \pdfbar{0.937} & Cst. & \pdfbar{0.019} \\
        W.B. & \pdfbar{0.271} & BW & \pdfbar{0.058} \\
        Satu. & \pdfbar{0.076} & Color & \pdfbar{0.021} \\
        \hline
    \end{tabular}
    }
};
\node (agent4d) at ([yshift=-2.0em]agent4.center)
{Tone\quad\quad\quad\quad};
\draw[<->] (\DeltaX+4.000000,\DeltaY-9.220000) -- (\DeltaX+4.000000,\DeltaY-10.100000) -- (\DeltaX+4.880000,\DeltaY-10.100000);
\draw[-] (\DeltaX+4.000000,\DeltaY-10.100000) -- (\DeltaX+4.100000,\DeltaY-9.985508);
\draw[-] (\DeltaX+4.100000,\DeltaY-9.985508) -- (\DeltaX+4.200000,\DeltaY-9.767321);
\draw[-] (\DeltaX+4.200000,\DeltaY-9.767321) -- (\DeltaX+4.300000,\DeltaY-9.584581);
\draw[-] (\DeltaX+4.300000,\DeltaY-9.584581) -- (\DeltaX+4.400000,\DeltaY-9.528366);
\draw[-] (\DeltaX+4.400000,\DeltaY-9.528366) -- (\DeltaX+4.500000,\DeltaY-9.473776);
\draw[-] (\DeltaX+4.500000,\DeltaY-9.473776) -- (\DeltaX+4.600000,\DeltaY-9.419177);
\draw[-] (\DeltaX+4.600000,\DeltaY-9.419177) -- (\DeltaX+4.700000,\DeltaY-9.364566);
\draw[-] (\DeltaX+4.700000,\DeltaY-9.364566) -- (\DeltaX+4.800000,\DeltaY-9.300000);
\node[draw, rectangle, thick,minimum height=7em,minimum width=7em](agent5) at (11.300000,-3.000000) {};
\node (agent5s) at ([yshift=1.4em]agent5.center) {
    \scalebox{0.7}{
    \begin{tabular}{|p{0.5cm}p{0.2cm}p{0.5cm}p{0.2cm}|}
        \hline
        Expo. & \pdfbar{0.019} & Tone & \pdfbar{0.019} \\
        Gam. & \pdfbarSelected{2.627} & Cst. & \pdfbar{0.019} \\
        W.B. & \pdfbar{0.156} & BW & \pdfbar{0.034} \\
        Satu. & \pdfbar{0.103} & Color & \pdfbar{0.021} \\
        \hline
    \end{tabular}
    }
};
\node (agent5d) at ([yshift=-2.0em]agent5.center)
{Gamma $1/0.77$};
        \foreach \from/\to in {image1/agent1, image2/agent2, image3/agent3, image4/agent4, image5/agent5}
            \draw[->,very thick] (\from.east) -- (\to.west);
        \foreach \from/\to in {agent1/image2, agent2/image3, agent4/image5}
            \draw[->,very thick] (\from.south) -- ++(0,-0.1) -- ++(0,-0.2) -| (\to.north);
        \draw[->,very thick] (agent3.east) -- ++(0.5,0.0) -- ++(-0.0,0.0) |- (image4.west);
        \draw[->,very thick] (agent5.east) -- ++(0.4,0.0) -- ++(-0.0,0.0) |- (final.west);
    \end{tikzpicture}
    }
    \caption{Our method provides automatic and end-to-end processing of RAW photos,
    directly from linear RGB data captured by camera sensors to visually pleasing and
    display-ready images. Our system not only generates appealing results, but also outputs a meaningful operation sequence.
    User studies indicate that the retouching results surpass those of strong baselines, even though our method uses only unpaired data for training.}
    \label{fig:teaser}
\end{teaserfigure}
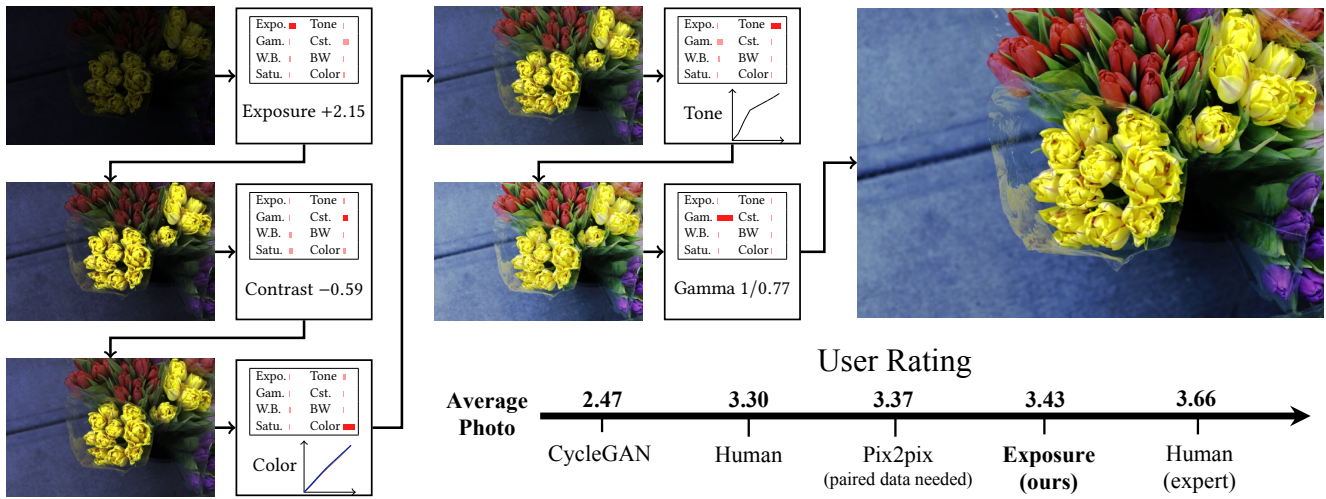
\if(0)
\begin{teaserfigure}
    \centering
    \includegraphics[width=1.0\linewidth]{images/teaser-draft}
    \caption{Our method provides automatic and end-to-end processing of RAW photos,
    directly from linear RGB data captured by camera sensors to visually pleasing and
    display-ready images. Our system not only generates appealing results, but also outputs a meaningful operation sequence.
    User studies indicate that the retouching results surpass those of strong baselines, even though our method uses only unpaired data for training.}
    \label{fig:teaser}
\end{teaserfigure}
\fi

\newcommand{\mdp}{\mathcal{P}}
\newcommand{\statespace}{\mathcal{S}}
\newcommand{\actionspace}{\mathcal{A}}
\newcommand{\policy}{\pi}
\newcommand{\rlReturn}{r^{\gamma}}
\newcommand{\grad}{\nabla}
\newcommand{\etal}{et al. }
\newcommand{\TODO}[1]{\textcolor{red}{[#1]}}
\newcommand{\E}{\mathop{\mathbb{E}}}
\newcommand{\changed}[2]{#2}
\newcommand{\Changed}[2]{#2}

\newcommand{\added}[1]{\textcolor{red}{#1}}
\newcommand{\placeholder}{\adjustbox{margin=1em,width=0.45\textwidth,set height=4cm,set depth=4cm,frame,center}{Placeholder}}
\newcommand{\folder}{1880}

\newlength{\mystrutht}
\newcommand{\mystrut}{\rule{0pt}{\mystrutht}\rule[-1pt]{0pt}{1pt}}
\newcolumntype{C}{>{\hbox to 1em\bgroup\mystrut\hfil}c<{\hfil\egroup}}
\newenvironment{zerotabular}
  {\settoheight{\mystrutht}{A}\addtolength{\mystrutht}{1pt}%
   \setlength{\tabcolsep}{0pt}%
   \renewcommand{\arraystretch}{0}%
   \begin{tabular}}
  {\end{tabular}}

\newcommand{\Boxed}[1]{{\setlength{\fboxsep}{0pt}\color{white}\setlength{\fboxrule}{0.7pt}\fbox{#1}\color{black}}}

\newcommand{\GG}[2]{\Boxed{\includegraphics[width=.238\textwidth]{images/gallery3/#1/#2}}}
\newcommand{\GGO}[2]{
\Boxed{\includegraphics[width=.238\linewidth]{images/inputs_tone_mapped/#2}} &
\Boxed{\includegraphics[width=.238\linewidth]{images/gallery3/#1/#2}}
}
\newcommand{\GGOH}[1]{
\Boxed{\includegraphics[width=.159\textwidth]{images/inputs_tone_mapped/#1}} &
\Boxed{\includegraphics[width=.159\textwidth]{images/hawaii/#1}}
}

\newcommand{\GGG}[1]{
\Boxed{\includegraphics[width=.155\textwidth]{images/inputs_tone_mapped/#1}}&
\Boxed{\includegraphics[width=.155\textwidth]{images/gallery4/fhpA/#1}}&
\Boxed{\includegraphics[width=.155\textwidth]{images/gallery4/fhpB/#1}}
}

\setlength{\tabcolsep}{4pt}

\begin{abstract}

Retouching can significantly elevate the visual appeal of photos, but many casual photographers lack the expertise to do this well. To address this problem, previous works have proposed automatic retouching systems based on supervised learning from {\em paired} training images acquired before and after manual editing. As it is difficult for users to acquire paired images that reflect their retouching preferences, we present in this paper
a deep learning approach that is instead trained on {\em unpaired} data, namely a set of photographs that exhibits a retouching style the user likes, which is much easier to collect.
Our system is formulated using deep convolutional neural networks that learn to apply different retouching operations on an input image. Network training with respect to various types of edits is enabled by modeling these retouching operations in a unified manner as resolution-independent differentiable filters. To apply the filters in a proper sequence and with suitable parameters, we employ a deep reinforcement learning approach that learns to make decisions on what action to take next, given the current state of the image. In contrast to many deep learning systems, ours provides users with an understandable solution in the form of conventional retouching edits, rather than just a \Changed{``black box''}{``black-box''} result. Through quantitative comparisons and user studies, we show that this technique generates retouching results consistent with the provided photo set.

\end{abstract}


%
%


%
%

\maketitle

\section{Introduction}

The aesthetic quality of digital photographs can be appreciably enhanced through retouching. Experienced photographers often perform a variety of post-processing edits, such as color adjustment and image cropping, to produce a result that is expressive and more visually appealing. Such edits can be applied with the help of software such as {\it Photoshop} and {\it Lightroom}; however, photo retouching remains challenging for ordinary users who lack the skill to manipulate their images effectively. This problem underscores the need for automatic photo editing tools, which can be helpful even to professionals by providing a better starting point for manual editing.

An important consideration in photo retouching is that different people have different preferences in retouching style. While some people like photos with vibrant colors that pop out of the screen, others may prefer more subdued and natural coloring, or even a monochromatic look.
Personal preferences extend well beyond color to include additional image properties such as contrast and tone.
They furthermore may vary with respect to semantic image content.

A natural way to express a user's personal preferences is through a set of retouched photos that they find appealing. The photographs may be self-curated from the web or taken from the collection of a favorite photographer. Such an image set gives examples of what an automatic retouching system should aim for,
but few techniques are able to take advantage of this guidance.
The automatic editing tools in the literature are mostly designed to handle only a single aspect of photo retouching, such as image cropping~\cite{yan2015change}, tonal adjustment~\cite{bychkovsky2011learning}, and color enhancement~\cite{wang2011example, yan2014learning, yan2016automatic}. Moreover, the state-of-the-art techniques for these problems are all based on machine learning with training data composed of image pairs, before and after the particular retouching operation. Paired image data is generally difficult to acquire, as images prior to retouching are usually unavailable. This is especially true of photos before and after a specific retouching step.

In this paper, we present a photo retouching system that handles a wide range of post-processing operations within a unified framework, and learns how to apply these operations based on a photo collection representing a user's personal preferences. No paired image data is needed. This is accomplished through an end-to-end learning framework in which various retouching operations are formulated as a series of resolution-independent differentiable filters that can be jointly trained within a convolutional neural network (CNN). How to determine the sequence and parameters of these filters for a given input image is learned with a deep reinforcement learning (RL) approach guided by a generative adversarial network (GAN) that models personal retouching preferences from a given photo collection.

\begin{figure}
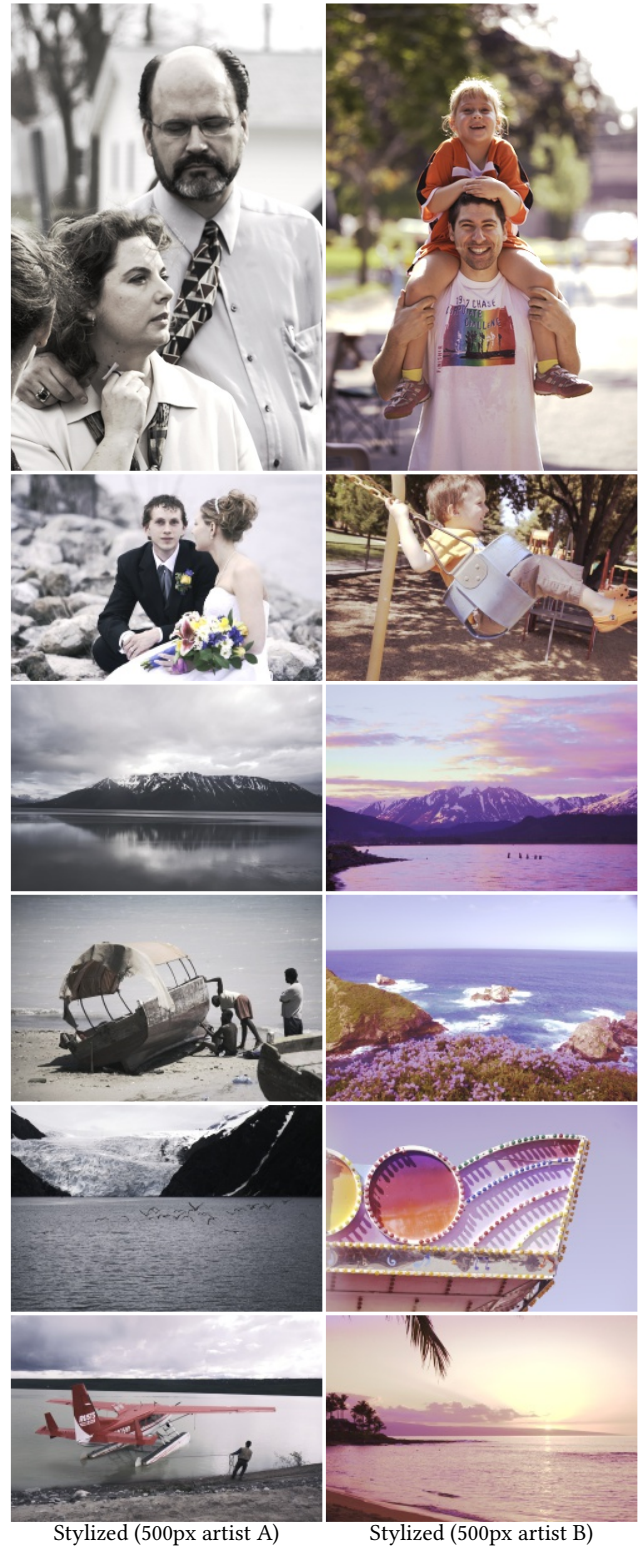

   \centering
   \scalebox{0.97}{
   \begin{zerotabular}{cc}
   \GG{fhpA}{2024} & \GG{fhpB}{1688} \\
   \GG{fhpA}{4082} & \GG{fhpB}{4871} \\
   \GG{fhpA}{3609} & \GG{fhpB}{1328} \\
   \GG{fhpA}{2597} & \GG{fhpB}{1651} \\
   \GG{fhpA}{2631} & \GG{fhpB}{3196} \\
   \GG{fhpA}{3654} & \GG{fhpB}{3913} \\
   Stylized (500px artist A) & Stylized (500px artist B)
   \end{zerotabular}
   }
  \caption{\changed{}{Starting from RAW inputs, our system can produce retouched results with styles learned from professional photographers.}}
  \label{fig:final2}
\end{figure}

In contrast to many neural network solutions where the functioning is hidden within a ``black box'', our ``white box'' network can reveal its sequence of editing steps for an image, which correspond to standard retouching operations and provide some understanding of the process that it took. We show both quantitatively and through user studies that the system can produce results that plausibly match the user preferences reflected within a photo collection. Examples of these automatically retouched images are displayed in Figure~\ref{fig:teaser} and Figure~\ref{fig:final2}.

The technical contributions of this work are summarized as follows:
\begin{itemize}
\item An end-to-end model of photo post-processing with a set of differentiable filters.
\item By optimizing the model using reinforcement learning, our system can generate a meaningful operation sequence that provides users with an understanding of the given artistic style, rather than just outputting a black-box result.
\item Using a GAN structure, we enable learning of photo retouching without image pairs. To our knowledge, this is the first GAN that scales with image resolution and generates no distortion artifacts in the image.
\item Through extensive experiments, we qualitatively and quantitatively validate our model and learning framework. We show that our method not only provides an effective end-to-end post-processing tool that aids ordinary users, but also can help advanced users to reverse-engineer the style of an automatic filter.
\end{itemize}

\section{Related Work}

\paragraph{Automatic Photo Retouching}

The state-of-the-art methods for automatic retouching are mainly based on supervised learning from paired images, which are obtained before and after editing by an expert photographer. Most of these methods extract handcrafted features, such as intensity distributions and scene brightness, from an input image and learn to determine editing parameters with respect to them. This approach has been employed for individual types of post-processing operations, including global tonal adjustment using Gaussian processes regression~\cite{bychkovsky2011learning}, image cropping using support vector machines~\cite{yan2015change,fang2014automatic}, and color adjustment using a learning-to-rank approach~\cite{yan2014learning} or with binary classification trees and parametric mapping models~\cite{wang2011example}.

With recent developments in machine learning, more informative features have been extracted from images using deep convolutional neural networks. In contrast to the low-level image properties represented by handcrafted features, the features from deep learning encode high-level semantic information, from which context-dependent edits can be learned. Deep CNNs have led to clear improvements in a wide range of computer graphics applications, including 3D mesh labeling~\cite{guo20153d} and locomotion modeling~\cite{peng2016terrain}. For photo retouching, CNNs have been utilized for spatially varying color mapping based on semantic information together with handcrafted global and local features~\cite{yan2016automatic}, and for color constancy~\cite{hu2017fc} 
where semantic understanding helps in resolving estimation ambiguity.
More recently, a CNN was trained to predict local affine transforms in bilateral space~\cite{gharbi2017deep}, which can serve as an approximation to edge-aware image filters and color/tone adjustments.

Our system utilizes deep CNNs as well, but differs from these previous works in that it generates a meaningful sequence of edits that can be understood and reproduced by users. Moreover, it performs learning {\it without} paired image data. Not only is
collecting unpaired data
more practical from a user's perspective, but we believe that it more closely conforms with the task of retouching. Unlike the one-to-one mapping that is implicit in supervised learning, retouching is inherently a one-to-many problem, since for a given input image there exist many possible solutions that are consistent with a retouching style. Instead of learning to convert a given photograph into a specific result, our technique learns to transform an image into a certain style as represented by a photo collection.

Related to \Changed{this}{our approach} is a method for finding exemplar images whose color and tone style is compatible with a given input photograph~\cite{lee2016automatic}. This is done by semantically matching the input to clusters of images from a large dataset, and then sampling from a set of stylized images that match these clusters in chrominance and luminance distributions. Matching through these two degrees of separation may lead to exemplar images that differ drastically from the input in content and composition ({\it e.g.}, a landscape photo as a style exemplar for a closeup image of a child). In such cases, mapping the chrominance distribution of the exemplar to the input may produce unusual colorings. By contrast, our system models a style from a collection of photos, and transforms the input image towards this style, rather than to the statistics of a particular image.

\changed{}{Image colorization (e.g., ~\cite{larsson2016learning, zhang2016colorful}) is \Changed{another}{a} problem related \Changed{}{to photo retouching}. However, apart from processing colors, our system is also capable of more general image editing operations, including adjustments to exposure, tone and contrast.}

\changed{}{
There exist software tools such as {\em Aperture} that can
automatically adjust photos. Notably, in {\em Aperture 3.3} such adjustment is done via multiple operations,
which are visible to the user and allow further adjustments to be done using its predefined operations.
Our system similarly provides transparency and further editing, while also learning the artistic style from a set of example images.
Moreover, it reveals the underlying actions of the deep neural networks, which are usually considered as black boxes.}

\paragraph{Data-Driven Image Generation}

Early methods for data-driven image generation addressed problems such as texture synthesis~\cite{efros1999texture} and super-resolution~\cite{freeman2002example} through sampling or matching of image patches from a database. To generate a certain class of images such as handwritten digits or human faces, there has been some success using variational auto-encoders~\cite{kingma2014auto,rezende2014stochastic}, which construct images from a compressed latent representation learned from a set of example images.

\paragraph{Generative Adversarial Networks}

Recently, significant progress in data-driven image generation has been achieved through generative adversarial networks (GANs) ~\cite{goodfellow2014generative}.
GANs are composed of two competing networks, namely a generator that learns to map from a latent space to a target data distribution ({\it e.g.}, natural images), and a discriminator that learns to distinguish between instances of the target distribution and the outputs of the generator. The generator aims to better mimic the target data distribution based on feedback from the discriminator, which likewise seeks to improve its discriminative performance. Through this adversarial process, GANs have generated images with a high level of realism~\cite{radford2016unsupervised}.

A conditional variant of GANs~\cite{mirza2014conditional} makes it possible to constrain the image generation using information in an input image. Conditional GANs have been applied to image inpainting conditioned on the surrounding image context~\cite{pathak2016context}, inferring photographic images from surface normal maps~\cite{wang2016generative}, super-resolution from a low-resolution input~\cite{ledig2016photo}, and image stylization for an input image and a texture example~\cite{li2016precomputed}. These image-to-image translation problems were modeled within a single framework under paired~\cite{isola2016image} and unpaired~\cite{zhu2017unpaired} settings. These two methods are based on the observation that GANs learn a loss function that adapts to the data, so they can be applied in the same way to different image-to-image translation tasks. For CycleGAN~\cite{zhu2017unpaired} with unpaired training data, the translations are encouraged to be ``cycle consistent'', where a translation by the GAN from one domain to another should be reverted back to the original input by another counterpart GAN. \changed{}{The Adversarial Inverse Graphics Network (AIGN)~\cite{tung2017adversarial}, by utilizing a problem-specific renderer, can make use of unpaired data for image-to-image translation as well.}
\changed{}{We refer the readers to a good survey ~\cite{wu2017survey} of GANs on image generation tasks.}

Our system also uses a type of conditional GAN, but instead of directly generating an image, it outputs the parameters of filters to be applied to the input image. As the filters are designed to be content-preserving, this approach maintains the semantic content and spatial structure of the original image, as is the case for CycleGAN. Also, since the filters are resolution-independent, they can be applied to images of arbitrary size ({\it e.g.}, 24-megapixel photos), even though GANs in practice can generate images of only limited resolution ({\it e.g.}, 512$\times$512px in CycleGAN). In addition, the generated filtering sequence represents conventional post-processing operations understandable to users, unlike the black-box solutions of most CNNs. Concurrently to our work, another deep learning based solution is proposed in \cite{fang2017creatism}, which also makes use of a conditional GAN structure, but only for the ``dramatic mask'' part of the system. In addition, multiple operations are learned separately, while in our work operations are optimized elegantly as a whole, guided by the RL and GAN architecture.

\paragraph{Reinforcement Learning}

Different from existing GAN architectures, our conditional GAN is incorporated within a reinforcement learning (RL) framework to learn a sequence of filter operations. RL provides models for \Changed{decision making}{decision-making} and agent interaction with the environment, and has led to human-level performance in playing Atari games~\cite{mnih2013playing} and even defeating top human competitors at the game of Go~\cite{silver2016mastering}. In graphics, RL has been successfully used for character animation~\cite{peng2015dynamic,peng2016terrain,peng2017learning,peng2017deeploco}. For natural language generation, a combination of RL and GAN was employed in~\cite{yu2017seqgan} so that sequences consisting of discrete tokens can be effectively produced. In our work, a filtering sequence is modeled as a series of \Changed{decision making}{decision-making} problems, with an image quality evaluator defined by the GAN discriminator as the environment.

\paragraph{On-camera enhancement of image quality}

For mobile phone cameras, methods have been developed to automatically enhance photos within the imaging pipeline. These enhancements have included image denoising~\cite{Liu2014fast} as well as exposure adjustment and tone mapping~\cite{hasinoff2016burst}. These techniques are aimed at improving generic image quality and do not address emulating retouching styles. Also, these methods operate on a set of burst images to obtain data useful for their tasks. Our work can be employed in conjunction with such methods, for further processing to improve or personalize the photographic style.

\section{The Model}

Given an input RAW photo, the goal of our work is to generate a result that is retouched to match a given photo collection. The photo collection may represent a specific photographic style, a particular photographer, or a more general range of image appearance, such as an assembled set of photos that the user likes. In this section, we elaborate on our modeling of the retouching process.

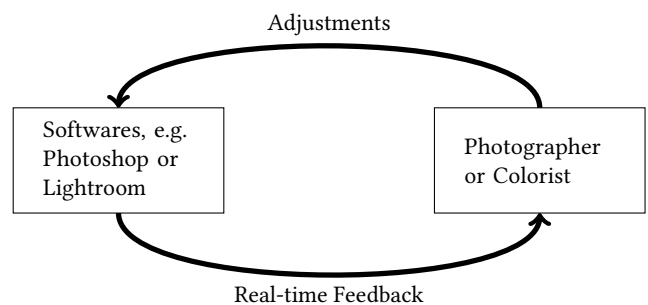
\begin{figure}[b]
  \centering
  \begin{tikzpicture}[scale=0.7]
    \draw [black] (0,0) rectangle (4, 2) node[pos=0.5, text width=2.0cm] {Softwares, e.g. Photoshop or Lightroom};
    \draw [black] (8,0) rectangle (12, 2) node[pos=0.5, text width=2.0cm] {Photographer or Colorist};
    \draw [line width=2pt, <-] (2,2) to [out=90, in=90, looseness=0.5,
          edge node = {node[pos=0.5,above]{Adjustments}}] (10,2);
    \draw [line width=2pt, ->] (2,0) to [out=-90, in=-90, looseness=0.5,
          edge node = {node[pos=0.5,below]{Real-time Feedback}}] (10,0);
  \end{tikzpicture}
  \caption{Information flow in interactive photo post-processing. Our method follows this scheme by modeling retouching as a decision-making sequence.}
  \label{fig:interactive}
\end{figure}

\subsection{Motivation}
In contrast to most methods that address a single post-processing operation, a more comprehensive retouching process needs to account for different types of edits and how to apply them collectively. For a human editor, retouching is done as a series of editing steps, where each step is normally decided based on the outcome of the previous step. This reliance on visual feedback exists even within a single step of the process, since the parameters for an operation, often controlled with a slider, are interactively adjusted while viewing real-time results, as shown in Figure~\ref{fig:interactive}.
\changed{}{Such step-wise modeling of image editing has led to many interesting works in HCI, such as~\cite{grabler2009generating, chen2011nonlinear, chen2016data}.}

Certainly, feedback is critical for choosing an operation and its parameters. A photographer cannot in general determine a full operation sequence from viewing only the original input image. We postulate that an automatic retouching system would also benefit from feedback and can more effectively learn how to select and apply a single operation at a time based on feedback than to infer the final output directly from the input. Moreover, modeling retouching as a sequence of standard post-processing operations helps to maintain the photorealism of the image and makes the automatic process more understandable to users.

We note that the notion of learning an operation sequence was used in a learning-to-rank model for automatic color adjustment~\cite{yan2014learning}. Unlike their supervised approach which is trained on collected sequences of editing operations from expert photographers, our system requires much less supervision, needing only a set of retouched photos for training.
\changed{}{In ~\cite{Hu13InverseImageEditing}, editing sequences are recovered given input and output image pairs based on region matching, and further adjustment of these sequences is made possible.}

\subsection{Post-processing as a decision-making sequence}
Based on this motivation, the retouching process can naturally be modeled as a sequential decision-making problem, which is a problem commonly addressed in reinforcement learning (RL). RL is a subarea of machine learning related to how an agent should act within an environment to maximize its cumulative rewards. Here, we briefly introduce basic concepts from RL and how we formulate retouching as an RL problem.

\begin{figure}
  \resizebox{1.03\linewidth}{!}{
    \centering
    \begin{tikzpicture}[scale=0.95]
        \node[inner sep=0pt] (image1) at (0,0) {\includegraphics[width=.19\textwidth]{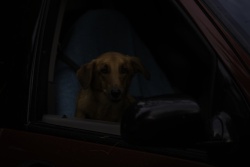}};
        \node[inner sep=0pt] (image2) at (0,-3) {\includegraphics[width=.19\textwidth]{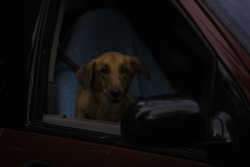}};
        \node[inner sep=0pt] (image3) at (0,-6) {\includegraphics[width=.19\textwidth]{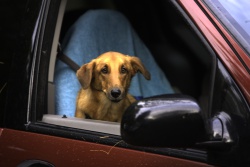}};
        \node[inner sep=0pt] (image4) at (0,-9) {\includegraphics[width=.19\textwidth]{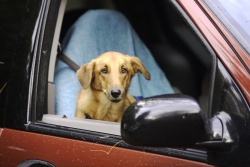}};
        \node[inner sep=0pt] (image5) at (0, -12) {\includegraphics[width=.19\textwidth]{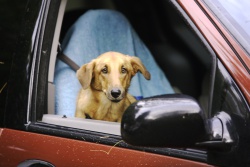}};
        \node[inner sep=0pt] (image6) at (1.75, -16.5) {\includegraphics[width=.372\textwidth]{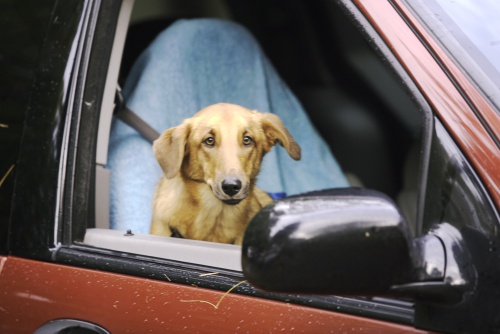}};
        \node(s0) at ([xshift=-1.0em]image1.west) {\LARGE{$s_0$}};
        \node(s1) at ([xshift=-1.0em]image2.west) {\LARGE{$s_1$}};
        \node(s2) at ([xshift=-1.0em]image3.west) {\LARGE{$s_2$}};
        \node(s3) at ([xshift=-1.0em]image4.west) {\LARGE{$s_3$}};
        \node(s4) at ([xshift=-1.0em]image5.west) {\LARGE{$s_4$}};
        \node(s5) at ([xshift=-1.0em]image6.west) {\LARGE{$s_5$}};

\definecolor{tempcolor}{RGB}{134.5,125.5,113}
\node[draw, rectangle, thick,minimum height=7em,minimum width=7em](agent1) at (4.000000,0.000000) {};
\node (agent1s) at ([yshift=1.4em]agent1.center) {
    \scalebox{0.7}{
    \begin{tabular}{|p{0.5cm}p{0.2cm}p{0.5cm}p{0.2cm}|}
        \hline
        Expo. & \pdfbar{0.531} & Tone & \pdfbar{0.120} \\
        Gam. & \pdfbarSelected{1.900} & Cst. & \pdfbar{0.094} \\
        W.B. & \pdfbar{0.058} & BW & \pdfbar{0.186} \\
        Satu. & \pdfbar{0.071} & Color & \pdfbar{0.041} \\
        \hline
    \end{tabular}
    }
};
\node (agent1d) at ([yshift=-2.0em]agent1.center)
{Gamma $1/3.00$};
\node[draw, rectangle, thick,minimum height=7em,minimum width=7em](agent2) at (4.000000,-3.000000) {};
\node (agent2s) at ([yshift=1.4em]agent2.center) {
    \scalebox{0.7}{
    \begin{tabular}{|p{0.5cm}p{0.2cm}p{0.5cm}p{0.2cm}|}
        \hline
        Expo. & \pdfbarSelected{2.467} & Tone & \pdfbar{0.124} \\
        Gam. & \pdfbar{0.020} & Cst. & \pdfbar{0.054} \\
        W.B. & \pdfbar{0.095} & BW & \pdfbar{0.112} \\
        Satu. & \pdfbar{0.071} & Color & \pdfbar{0.057} \\
        \hline
    \end{tabular}
    }
};
\node (agent2d) at ([yshift=-2.0em]agent2.center)
{Exposure $+0.41$};
\node[draw, rectangle, thick,minimum height=7em,minimum width=7em](agent3) at (4.000000,-6.000000) {};
\node (agent3s) at ([yshift=1.4em]agent3.center) {
    \scalebox{0.7}{
    \begin{tabular}{|p{0.5cm}p{0.2cm}p{0.5cm}p{0.2cm}|}
        \hline
        Expo. & \pdfbar{0.020} & Tone & \pdfbar{0.569} \\
        Gam. & \pdfbar{0.019} & Cst. & \pdfbar{0.416} \\
        W.B. & \pdfbar{0.560} & BW & \pdfbar{0.173} \\
        Satu. & \pdfbar{0.256} & Color & \pdfbarSelected{0.989} \\
        \hline
    \end{tabular}
    }
};
\node (agent3d) at ([yshift=-2.0em]agent3.center)
{Color\quad\quad\quad\quad};
\draw[<->] (4.000000,-6.220000) -- (4.000000,-7.100000) -- (4.880000,-7.100000);
\draw[red,-] (4.000000,-7.100000) -- (4.100000,-7.007718);
\draw[red,-] (4.100000,-7.007718) -- (4.200000,-6.915392);
\draw[red,-] (4.200000,-6.915392) -- (4.300000,-6.822582);
\draw[red,-] (4.300000,-6.822582) -- (4.400000,-6.709906);
\draw[red,-] (4.400000,-6.709906) -- (4.500000,-6.597244);
\draw[red,-] (4.500000,-6.597244) -- (4.600000,-6.484737);
\draw[red,-] (4.600000,-6.484737) -- (4.700000,-6.392244);
\draw[red,-] (4.700000,-6.392244) -- (4.800000,-6.300000);\draw[green,-] (4.000000,-7.100000) -- (4.100000,-7.007061);
\draw[green,-] (4.100000,-7.007061) -- (4.200000,-6.914088);
\draw[green,-] (4.200000,-6.914088) -- (4.300000,-6.819288);
\draw[green,-] (4.300000,-6.819288) -- (4.400000,-6.706237);
\draw[green,-] (4.400000,-6.706237) -- (4.500000,-6.593326);
\draw[green,-] (4.500000,-6.593326) -- (4.600000,-6.494879);
\draw[green,-] (4.600000,-6.494879) -- (4.700000,-6.393101);
\draw[green,-] (4.700000,-6.393101) -- (4.800000,-6.300000);\draw[blue,-] (4.000000,-7.100000) -- (4.100000,-6.987723);
\draw[blue,-] (4.100000,-6.987723) -- (4.200000,-6.875444);
\draw[blue,-] (4.200000,-6.875444) -- (4.300000,-6.783379);
\draw[blue,-] (4.300000,-6.783379) -- (4.400000,-6.691486);
\draw[blue,-] (4.400000,-6.691486) -- (4.500000,-6.599596);
\draw[blue,-] (4.500000,-6.599596) -- (4.600000,-6.507654);
\draw[blue,-] (4.600000,-6.507654) -- (4.700000,-6.412237);
\draw[blue,-] (4.700000,-6.412237) -- (4.800000,-6.300000);
\node[draw, rectangle, thick,minimum height=7em,minimum width=7em](agent4) at (4.000000,-9.000000) {};
\node (agent4s) at ([yshift=1.4em]agent4.center) {
    \scalebox{0.7}{
    \begin{tabular}{|p{0.5cm}p{0.2cm}p{0.5cm}p{0.2cm}|}
        \hline
        Expo. & \pdfbar{0.019} & Tone & \pdfbar{0.413} \\
        Gam. & \pdfbar{0.019} & Cst. & \pdfbar{0.190} \\
        W.B. & \pdfbarSelected{2.145} & BW & \pdfbar{0.097} \\
        Satu. & \pdfbar{0.094} & Color & \pdfbar{0.023} \\
        \hline
    \end{tabular}
    }
};
\node (agent4d) at ([yshift=-2.0em]agent4.center)
{\tikz \fill[tempcolor] (0,0) rectangle (8 ex, 2 ex);};
\node[draw, rectangle, thick,minimum height=7em,minimum width=7em](agent5) at (4.000000,-12.000000) {};
\node (agent5s) at ([yshift=1.4em]agent5.center) {
    \scalebox{0.7}{
    \begin{tabular}{|p{0.5cm}p{0.2cm}p{0.5cm}p{0.2cm}|}
        \hline
        Expo. & \pdfbar{0.019} & Tone & \pdfbarSelected{1.552} \\
        Gam. & \pdfbar{0.019} & Cst. & \pdfbar{1.026} \\
        W.B. & \pdfbar{0.020} & BW & \pdfbar{0.111} \\
        Satu. & \pdfbar{0.230} & Color & \pdfbar{0.024} \\
        \hline
    \end{tabular}
    }
};
\node (agent5d) at ([yshift=-2.0em]agent5.center)
{Tone\quad\quad\quad\quad};
\draw[<->] (4.000000,-12.220000) -- (4.000000,-13.100000) -- (4.880000,-13.100000);
\draw[-] (4.000000,-13.100000) -- (4.100000,-12.925429);
\draw[-] (4.100000,-12.925429) -- (4.200000,-12.863714);
\draw[-] (4.200000,-12.863714) -- (4.300000,-12.811631);
\draw[-] (4.300000,-12.811631) -- (4.400000,-12.727810);
\draw[-] (4.400000,-12.727810) -- (4.500000,-12.577225);
\draw[-] (4.500000,-12.577225) -- (4.600000,-12.505746);
\draw[-] (4.600000,-12.505746) -- (4.700000,-12.422901);
\draw[-] (4.700000,-12.422901) -- (4.800000,-12.300000);

        \node(a0) at ([xshift=1.0em]agent1.east) {\LARGE{$a_0$}};
        \node(a1) at ([xshift=1.0em]agent2.east) {\LARGE{$a_1$}};
        \node(a2) at ([xshift=1.0em]agent3.east) {\LARGE{$a_2$}};
        \node(a3) at ([xshift=1.0em]agent4.east) {\LARGE{$a_3$}};
        \node(a4) at ([xshift=1.0em,yshift=-9.5em]agent4.east) {\LARGE{$a_4$}};
        \foreach \from/\to in {image1/agent1, image2/agent2, image3/agent3, image4/agent4, image5/agent5}
            \draw[->,very thick] (\from.east) -- (\to.west);
        \foreach \from/\to in {agent1/image2, agent2/image3, agent3/image4, agent4/image5, agent5/image6}
            \draw[->,very thick] (\from.south) -- ++(0,-0.1) -- ++(0,-0.2) -| (\to.north);
    \end{tikzpicture}
    }
  \caption{An example trajectory of states (images) and actions (operations).}
  \label{fig:trajectory}
\end{figure}

We denote the problem as $\mdp=(\statespace, \actionspace)$ with $\statespace$ being the \textbf{state space} and $\actionspace$ the \textbf{action space}.
Specifically in our task, $\statespace$ is the space of images, which includes the RAW input image and all intermediate results in the automatic process, while $\actionspace$ is the set of all filter operations.
A \textbf{transition function} $p: \statespace \times \actionspace \rightarrow \statespace$ maps input state $s \in \statespace$ to its outcome state $s' \in \statespace$ after taking action $a\in \actionspace$. State transitions can be expressed as $s_{i+1}=p(s_i, a_i)$.
Applying a sequence of filters to the input RAW image results in a \textbf{trajectory} of states and actions:
$$t=(s_0, a_0, s_1, a_1, \ldots, s_{N-1}, a_{N-1}, s_N)$$
where $s_i \in \statespace$, $a_i \in \actionspace$ are states and actions,  $N$ is the number of actions, and $s_N$ is the stopping state, as shown in Figure~\ref{fig:trajectory}.
A central element of RL is the \textbf{reward function}, $r: \statespace \times \actionspace \rightarrow \mathbb{R}$, which evaluates actions given the state. Our goal is to select a \textbf{policy} $\policy$ that maximizes the accumulated reward during the decision-making process.
For this, we use a \textbf{stochastic policy} agent, where the policy $\policy: \statespace \rightarrow \mathbb{P}(\actionspace)$ maps the current state $s\in \statespace$ to $\mathbb{P}(\actionspace)$, the set of probability density functions over the actions. When an agent enters a state, it samples one action according to the probability density functions, receives the reward, and follows the transition function to the next state.

\renewcommand{\folder}{0694}

Given a trajectory $t=(s_0, a_0, s_1, a_1, \ldots, s_N)$, we define the \textbf{return} $\rlReturn_k$ as the summation of discounted rewards after $s_k$:
\begin{equation}
\rlReturn_k=\sum_{k'=0}^{N-k}\gamma^{k'}r(s_{k+k'},a_{k+k'}),
\end{equation}
where $\gamma\in [0, 1]$ is a \textbf{discount factor} which places greater importance on rewards in the nearer future. To evaluate a policy, we define the \textbf{objective}
\begin{equation}
J(\pi)=\E_{\substack{s_0\sim \statespace_0\\t\sim \policy}}\left[\rlReturn_0|\pi\right],
\end{equation}
where $s_0$ is the input image, \changed{}{$\E$ stands for expectation}, and $\statespace_0$ is the input dataset. Intuitively, the objective describes the expected return over all possible trajectories induced by the policy $\policy$.
The goal of the agent is to maximize the objective $J(\policy)$, which is related to the final image quality by the reward function $r$, as images (states) with high quality are more greatly rewarded.

The expected total discounted rewards on states and state-action pairs are defined by \textbf{state-value functions} $V$, and \textbf{action-value functions} $Q$:
\begin{eqnarray}
\label{eqn:V}V^\policy(s)&=&\E_{\substack{{s_0=s}\\{t\sim\policy}}}\left[\rlReturn_0\right]\\
\label{eqn:Q}Q^\policy(s,a)&=&\E_{\substack{s_0=s\\a_0=a\\t\sim\policy}}\left[\rlReturn_0\right].
\end{eqnarray}

To fit our problem into this RL framework, we decompose actions into two parts: a discrete selection of filter $a_1$ and a continuous decision on filter parameters $a_2$.
The policy also consists of two parts: $\policy=(\policy_1, \policy_2)$. $\policy_1$ is a function that takes a state and returns a probability distribution over filters, {\it i.e.} choices of $a_1$; and $\policy_2$ is a function that takes $(s, a_1)$ and then directly generates $a_2$.
Note that $\policy_1$ is stochastic and requires sampling. Since there are practical challenges in sampling a continuous random variable, we follow recent practices by treating $\policy_2$ deterministically, as described in section~\ref{policygradient}.

\section{Filter Design}
\label{sec:filter}
In this section, we discuss the design of filters, {\it i.e.} the action space $\actionspace$ in our model.

\begin{figure}[b]
   \begin{tikzpicture}[scale=1.1]
        \node[inner sep=0pt] (before_hi) at (0,0) {\includegraphics[width=.1\textwidth]{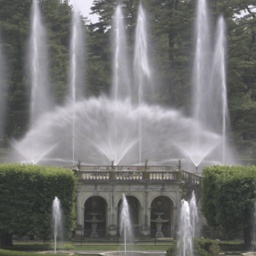}};
        \node[inner sep=0pt] (before_lo) at (3,0) {\includegraphics[width=.1\textwidth]{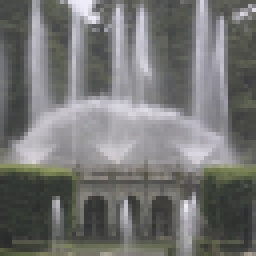}};
        \node[inner sep=0pt] (after_hi) at (0,-6) {\includegraphics[width=.1\textwidth]{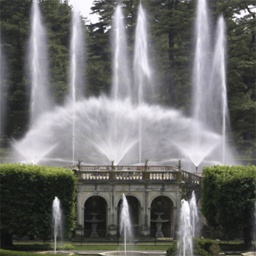}};
        \node[inner sep=0pt] (after_lo) at (3,-6) {\includegraphics[width=.1\textwidth]{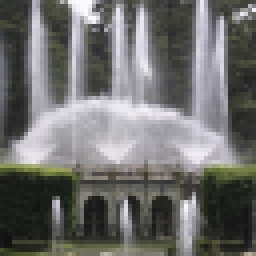}};

        \node[draw, rectangle, thick,minimum height=5.8em,minimum width=5.8em](nn) at (6.000000,0.000000) {};
        \node (nn_label1) at ([yshift=0.7em]nn.center) {
            Neural
        };
        \node (nn_label2) at ([yshift=-0.7em]nn.center) {
            Network
        };

        \node[draw, rectangle, thick,minimum height=7em,minimum width=7em](agent1) at (3.000000,-3.000000) {};
        \node (agent1s) at ([yshift=1.4em]agent1.center) {
            \scalebox{0.7}{
            \begin{tabular}{|p{0.5cm}p{0.2cm}p{0.5cm}p{0.2cm}|}
                \hline
                Expo. & \pdfbar{0.03} & W.B. & \pdfbar{0.02} \\
                Gam. & \pdfbar{0.2} & Satu. & \pdfbar{0.01} \\
                Color & \pdfbar{0.05} & Tone & \pdfbar{0.1} \\
                Level & \pdfbar{0.1} & Cst. & \pdfbarSelected{1.8} \\
                \hline
            \end{tabular}
            }
        };

        \node (before_hi_label) at ([yshift=1.0em]before_hi.north)
{High-res Input};

        \node (before_lo_label) at ([yshift=1.0em]before_lo.north)
{Low-res Input};

        \node (after_hi_label) at ([yshift=-1.0em]after_hi.south)
{High-res Output};

        \node (after_lo_label) at ([yshift=-1.0em]after_lo.south)
{Low-res Output};

        \node (agent1d) at ([yshift=-2.0em]agent1.center)
{Contrast $+0.95$};

        \draw[->,thick, dotted] (before_hi.east) -- (before_lo.west);

        \draw[->,thick] (before_lo.east) -- (nn.west);

        \draw[->,very thick] ([xshift=-0.3 em]nn.south) |- ([yshift=0.3 em]agent1.east) node[midway,sloped,above,rotate=0, label={[xshift=-2.5em,yshift=-0.9em]action}] {};

        \draw[->,very thick, red] ([yshift=-0.3 em]agent1.east) -| ([xshift=0.3 em]nn.south)
        node[midway,sloped,above,rotate=0, label={[xshift=-0.2em,yshift=-0.7em]gradient}] {};

        \draw[->,thick] (before_hi.south) to  [out=-90,in=180]([yshift=0.5 em]agent1.west) ;

        \draw[->,thick] ([yshift=-0.5 em]agent1.west) to [out=180,in=90] (after_hi.north) ;

        \draw[->,very thick] (before_lo.south) -- (agent1.north);

        \draw[->,very thick] ([xshift=-0.3 em] agent1.south) -- ([xshift=-0.3 em]after_lo.north) node[midway,sloped,above,rotate=180] {img.};
        \draw[->,very thick, red] ([xshift=0.3 em]after_lo.north) -- ([xshift=0.3 em]agent1.south) node[midway,sloped,above,rotate=180] {grad.};
    \end{tikzpicture}
  \caption{The design principles of our filters ({\em Contrast filter} in this example): 1) They are \textbf{differentiable} and can thereby provide gradients for neural network training; 2) \textbf{Arbitrary-resolution} images can be processed with the filter; 3) What the filter does should be \textbf{understandable} by a human user.}
  \label{fig:design-principles}
\end{figure}

\subsection{Design Principles}
For our system, we require the designs to adhere to the following properties.

\paragraph{Differentiable}
For gradient-based optimization of the policy $\policy$, the filters need to be differentiable with respect to their filter parameters. This differentiability is needed to allow training of the CNN by backpropagation. Clearly, not all filters can be trivially modeled as basic neural network layers; therefore, we propose approximations of such filters, such as piecewise linear functions in place of smooth curves, to incorporate them within our framework.

\paragraph{Resolution-independent}
Modern digital sensors capture RAW images at a high resolution ({\it e.g.}, $6,000\times 4,000$px) that is computationally impractical for CNN processing. Fortunately, most editing adjustments can be determined without examining an image at such high resolutions, thus allowing us to operate on downsampled versions of the RAW images. Specifically, we determine filter parameters on a low-resolution ($64\times 64$) version of a RAW image and then apply the same filter on the original high-resolution image. This strategy is similar to that used by Gharbi et al.~\shortcite{gharbi2017deep} to reduce computation on mobile devices. To this end, the filters need to be resolution-independent.

Note that most GAN-based image generation techniques, like CycleGAN~\cite{zhu2017unpaired}, generate images of resolution at around $512 \times 512$px, since higher resolutions lead to not only greater computational costs but also a significantly more challenging learning task that requires greater training time and training data. We experimentally compare our model to CycleGAN in section \ref{sec:cyclegan}.

\paragraph{Understandable}
The filters should represent operations that have an intuitive meaning, so that the generated operation sequence can be understood by users. This would be more interesting and instructive to users than a ``black-box'' result. It would also enable them to further adjust the parameters if they want. Though an alternative is to generate a black-box result and then let users apply edits to it, the black-box  transformation might not be invertible, leaving users unable to undo any unwanted effects.

\vspace*{0.1cm}
These three design principles are illustrated in Figure~\ref{fig:design-principles}.

\newcommand{\filterImgSize}{0.08\textwidth}
\newcommand{\insertFilter}[4]{\node[inner sep=0pt,
] (output#2) at (#1,-3) {\fbox{\includegraphics[width=\filterImgSize]{images/filters/output#2#3}}};
        \node[inner sep=0pt,
        ] (grad#2) at (#1,-5) {\fbox{\includegraphics[width=\filterImgSize]{images/filters/grad#2#3}}};
        \node(cap#2) [text width=3cm, align=center] at ([yshift=1.8em]output#2.north) {\small #4};}
\begin{figure*}
\hspace{-0.5cm}
\resizebox{.75\linewidth}{!}{
    \setlength{\fboxsep}{0pt}
    \setlength{\fboxrule}{0.5pt}
    \begin{tikzpicture}[scale=0.95]
        \node[inner sep=0pt,
        ] (image1) at (-1,0) {\fbox{\includegraphics[width=\filterImgSize]{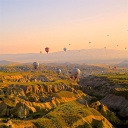}}};
        \node(inputCap) at ([yshift=1.0em]image1.north) {Input};
        \insertFilter{-8}{00}{E}{Exposure\\$+0.5$}
        \insertFilter{-6}{01}{G}{Gamma $2$}
        \insertFilter{-4}{02}{C}{Color Curve\\(Boost Red)}
        \insertFilter{-2}{03}{BW}{Black \& White \\$+0.5$}
        \insertFilter{0}{04}{W}{White Blanace\\(Blue)}
        \insertFilter{2}{05}{S+}{Saturaion\\$+0.5$}
        \insertFilter{4}{06}{T}{Tone Curve}
        \insertFilter{6}{07}{Ct}{Contrast $+0.8$}

        \node(outputCap) at ([xshift=-2.7em]output00.west) {Output};
        \node(gradCap) at ([xshift=-3.0em]grad00.west) {Gradient};

    \end{tikzpicture}
}
\hspace{-0.25cm}
\color{gray}
\hfill\vline\hfill
\color{black}
\hspace{0.25cm}
\includegraphics[width=0.25\textwidth]{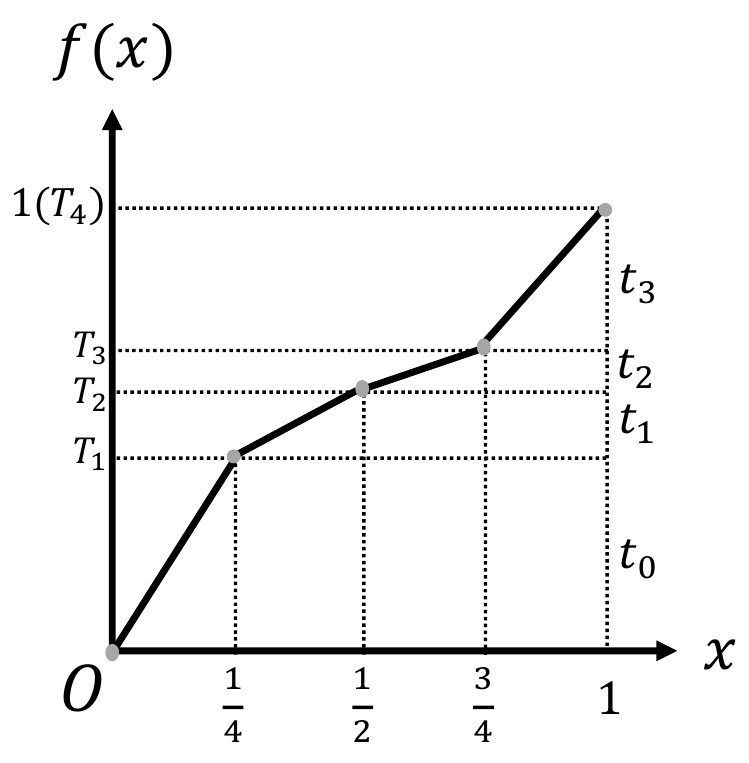}
\caption{\textbf{Left:} Visualizations of the eight differentiable filters. Gradients are displayed with a $+0.5$ offset so that negative values can be properly viewed. For the white balance filter, we visualize the gradient with respect to the red channel parameter; for the tone/color curves, we differentiate with respect to the first parameter of the \Changed{curve/red curve}{(red) curve}. \textbf{Right:} Representation of a curve using regressed parameters.}
\label{fig:filterGradCurve}
\end{figure*}

\begin{table}%
\caption{Simple pixel-wise filters. For more details about the filters, please refer to the supplemental material.}
\label{tab:simplefilters}
\hspace{-11mm}
\scalebox{0.88}{
\begin{minipage}{\columnwidth}
\begin{tabular}{lll}
  \toprule
  Operation & Parameters & Filter\\
  \hline
  Exposure              &              $E$: exposure value &   $p_O=2^E p_I$                    \\
  White Balance         &  $W_r, W_g, W_b$: factors & $p_O=(W_r r_I, W_g g_I, W_b b_I)$  \\
  Color curves          &  $C_{i,k}$: curve param. & $p_O=(L_{C_r}(r_I), L_{C_g}(g_I), L_{C_b}(b_I))$  \\
  \bottomrule
\end{tabular}
\bigskip\centering
\end{minipage}
}
\end{table}%

\subsection{Filter Details}
Based on the aforementioned design principles, we developed filters that map an input pixel value $p_I=(r_I, g_I, b_I)$ to an output pixel value $p_O=(r_O, y_O, g_O)$. Standard color and tone modifications, such as exposure change, white balancing, and color curve adjustment, can be modeled by such pixel-wise mapping functions. Examples of operations implemented in our system are listed in Table~\ref{tab:simplefilters} and visualized in Figure~\ref{fig:filterGradCurve} (left). Color curve adjustment, i.e., a channel-independent monotonic mapping function, requires special treatment for its filter to be differentiable, as described in the following.

\paragraph{Curve representation}

We approximate curves as monotonic \Changed{}{and} piecewise-linear functions, as illustrated in Figure~\ref{fig:filterGradCurve} (right). Suppose we represent a curve using $L$ parameters, denoted as $\{t_0, t_1, \ldots, t_{L-1}\}$. With the prefix-sum of parameters defined as
$T_k=\sum_{l=0}^{k - 1} t_l$, the points on the curves are represented as $(k/L, T_k/T_{L})$. For this representation, an input intensity $x\in [0, 1]$ will be mapped to
\begin{equation}
f(x)=\frac{1}{T_L}\sum_{i=0}^{L-1}\text{clip}(L\cdot x - i, 0, 1)t_k.
\end{equation}

Note that this mapping function is now represented by differentiable parameters, making the function differentiable with respect to both $x$ and the parameters $\{t_l\}$. Since color adjustments by professionals are commonly subtle, color curves are typically close to identity. We find that eight linear segments are sufficient for modeling typical color curves.

\section{Learning}

Given the decision-making model for retouching and the differentiable filters that make optimization possible, we discuss in this section how the agent is represented by deep neural networks (DNNs), how these networks are trained, and how the reward related to the generated image quality is evaluated using adversarial learning. The whole training cycle is shown in Alg.~\ref{alg:train}, and we elaborate on the details in the following subsections.

\subsection{Function approximation using DNNs}
DNNs are commonly deployed as an end-to-end solution for approximating functions used in complex learning tasks with plentiful data. Since convolutional neural networks (CNN) are especially powerful in image-based understanding~\cite{krizhevsky2012imagenet}, we use CNNs in our work.
Among the CNNs are two policy networks, which map the images into action probabilities $\policy_1$ (after softmax) or filter parameters $\policy_2$ (after $\tanh$). For policies $\policy_1$ and $\policy_2$, the network parameters are denoted as $\theta_1$ and $\theta_2$, respectively, and we wish to optimize $\theta=(\theta_1, \theta_2)$ so that the objective $J(\policy_{\theta})$ is maximized. In addition to the two policy networks, we also learn a value network and a discriminator network, which facilitate training as later described.

\begin{figure}
  \includegraphics[width=0.48\textwidth]{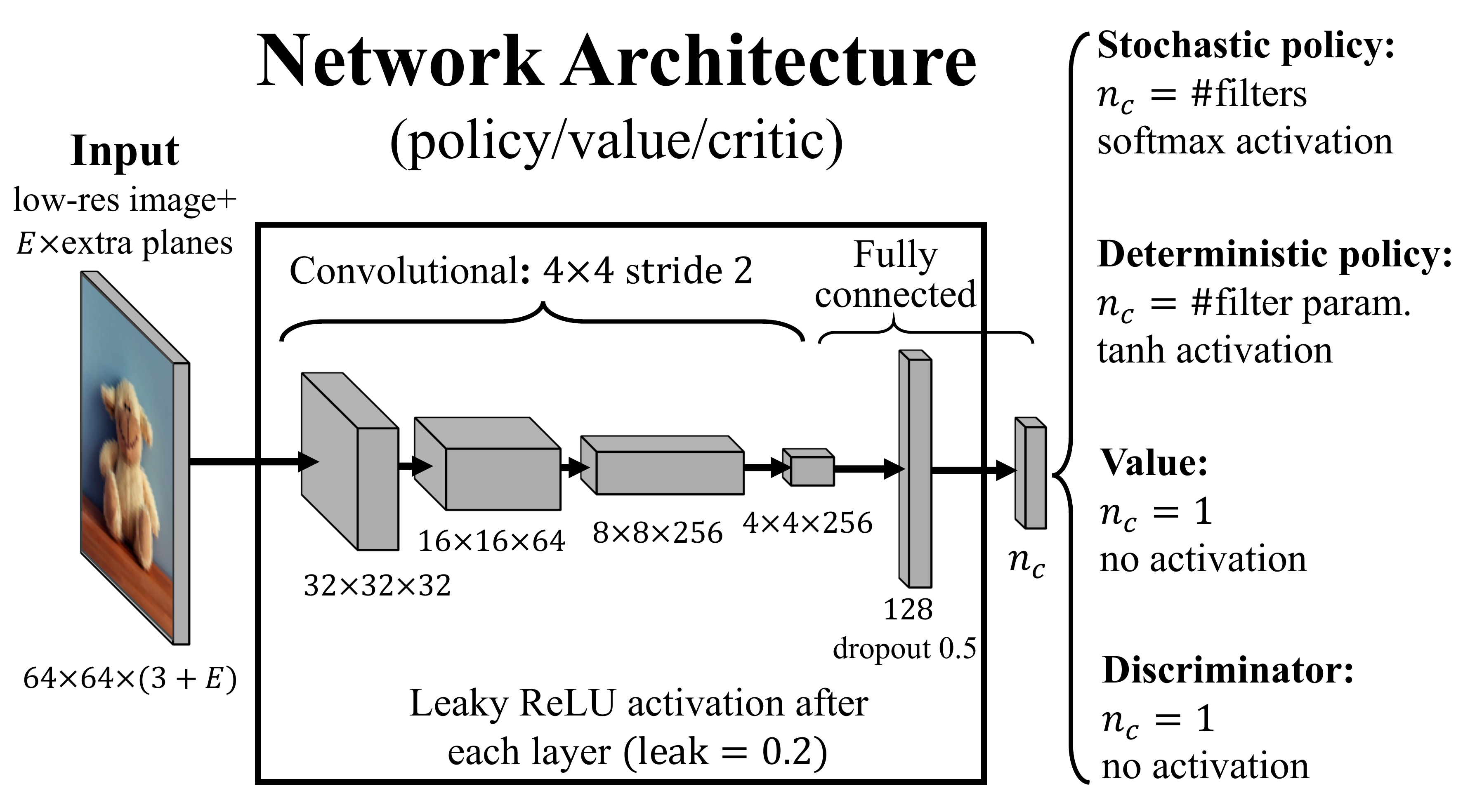}
  \caption{The general network structure shared by all networks in our system.}
  \label{fig:network}
\end{figure}

All of these networks share basically the same architecture illustrated in Figure~\ref{fig:network}, while having different numbers of output neurons according to what they output.
For each CNN, we use four convolution layers, each with filters of size $4\times 4$ and stride $2$. Appended to this is a fully connected layer to reduce the number of outputs to $128$, and then a final fully connected layer that further regresses the features into parameters we need from each network. \changed{}{After the first fully connected layer we apply dropout ($50\%$, during both training and test time) to provide noise to the generator, following~\cite{isola2016image}}. The deterministic policy networks (one for each filter) for filter parameter estimation share the convolutional layers, so that the computation is made more efficient. CNNs are largely tailored for hierarchical visual recognition, and we found that naively using them results in unsatisfactory learning of agent policies and global statistics. Therefore, following~\cite{silver2016mastering}, we concatenate extra (spatially constant) feature planes as additional color channels in the input. For the discriminator network, the additional feature planes are for the average luminance, contrast and saturation of the entire image; for the policy and value networks, the feature planes are eight boolean (zero or one) values that indicate which filters have been used, and another plane denotes the number of steps that have been taken so far in the retouching process.

\subsection{Policy network training}
\label{policygradient}
The policy networks are trained using policy gradient methods, which employ gradient descent to optimize parameterized policies with respect to the expected return. As the policy $\policy$ consists of two parts $(\policy_1, \policy_2)$ corresponding to the two decision-making steps ({\it i.e.}, filter and parameter selection), they are learned in an interleaved manner.

For filter selection, we sample $\policy_1$, which is a discrete probability distribution function $\policy_1(F_k)=\mathbb{P}[a_1=F_k]$ over all choices of filters $\mathcal{F}=\{F_1, F_2, \ldots, F_n\}$. Unlike other common differentiable operations including convolution or activation, the partial derivative $\partial{J(\policy)}/\partial \policy(F_k)$ cannot be directly calculated, which presents a challenge for backpropagation. We address this by applying the policy gradient theorem~\cite{sutton2000policy} to obtain an unbiased Monte Carlo estimate of the gradient of $J(\policy)$ with respect to $\policy_1$. For filter parameter selection, policy $\policy_2$ is deterministic, so that it is easier to optimize in a continuous space, and we formulate its gradient using the deterministic policy gradient theorem~\cite{silver2014deterministic}.
The policy gradients are thus expressed as
\begin{eqnarray}
\label{eqn:spg}\grad_{\theta_1}{J(\policy_{\theta})}&=&\E_{\substack{s\sim \rho^{\policy}\\a_1\sim\policy_1(s)\\a_2=\policy_{2}(s,a_1)}}[\grad_{\theta_1} \log \policy_1(a_1|s) Q(s,(a_1, a_2))],\\
\label{eqn:dpg}\grad_{\theta_2}{J(\policy_{\theta})}&=&\E_{\substack{s\sim \rho^{\policy}\\a_2=\policy_{2}(s,a_1)}}[\grad_{a_2}Q\left(s,(a_1, a_2)\right)\grad_{\theta_2} \policy_2(s, a_1) ],
\end{eqnarray}
where $\rho^{\pi}$ is the discounted state distribution defined as
\begin{equation}
\rho^{\pi}(s)=\sum_{\substack{k=0\\t\sim \pi}}^{\infty}\mathbb{P}(s_k=s)\gamma^k,
\end{equation}
and $Q$ is the value function defined in Eqn.~\ref{eqn:Q}.

To \Changed{calculate}{compute} these gradients, we apply the actor-critic \Changed{algorithm}{framework}~\cite{sutton2000policy}, where the actor is represented by the policy networks and the critic is the value network, which \Changed{learns the state-value function $V^{\pi}_{\nu}:\statespace\rightarrow \mathbb{R}$}{learns to approximate the state-value function $V^{\pi}$ (Eqn.~\ref{eqn:V}) using a CNN parameterized by $\nu$} \Changed{for estimating expected returns}{}. With the critic, the action-value function $Q^\policy$ can be computed by unfolding its definition (Eqn.~\ref{eqn:Q}) and expressing it in terms of the state-value function:
\begin{equation}
Q^\policy(s,a)=\E_{\substack{s_0=s\\a_0=a\\t\sim\policy}}[r(s_0, a_0) + \gamma V^{\policy}(p(s_0, a_0))].
\end{equation}
Plugging this into Eqn.~\ref{eqn:dpg} gives us the supervision signal for learning $\policy_2$.

We optimize the value network by minimizing
\begin{equation}
L_v=\E_{\substack{s\sim\rho^\policy\\a\sim\policy(s)}}\left[\frac{1}{2}\delta^2\right],
\end{equation}
where $\delta$ is the temporal difference (TD) error:
$\delta = r(s,a)+\gamma V\left(p(s,a)\right) - V(s).$
Note that $\delta$ \Changed{}{also} represents the Monte Carlo estimate of the \textbf{advantage} $A(s, a)=Q(s,a)-V(s)$, i.e., how much the value of action $a$ exceeds the expected value of actions at state $s$. For calculating the gradient of $\policy_1$ in Eqn.~\ref{eqn:spg}, the Q-value $Q(s,a)$ can be substituted by the advantage $A(s,a)$, which effectively reduces sample variance and can conveniently be computed as the TD error $\delta$. Note that the gradient of $\pi_2$ requires no Monte Carlo estimation, thus we directly calculate it by applying the chain rule on the gradient of $Q$, instead of using the advantage $A$.

\paragraph{Reward and discount factor} The ultimate goal is to obtain the best {\em final} result after all operations. For this, we set the reward as the incremental improvement in the quality score (modeled by a discriminator network in the following subsection) plus penalty terms (described in Sec.~\ref{subsec:penalizing}). We set the discount factor as $\gamma=1$ and allow the agent to make five edits to the input image. This number of edits was chosen to balance expressiveness and succinctness of the operation sequence.
\changed{}{We use a fixed number of steps because doing so makes training more stable than
having the network learn when to stop itself.
}

\begin{algorithm}[t]
\SetAlgoNoLine
\KwIn{Input datasets $D_{\text{RAW}}$ and $D_{\text{retouched}}$; batch size $b=64$, learning rates $\alpha_\theta=1.5\times 10^{-5}, \alpha_\omega=5\times 10^{-5}, \alpha_\nu=5\times 10^{-4}$, $n_{\text{critic}}=5$}
\KwOut{Actor model $\theta=(\theta_1, \theta_2)$, critic model $v$, and discriminator model $w$}
Initialize the trajectory buffer with $2,048$ RAW images\;
\While{$\theta$ has not converged}{
    \For{$i$ in $1..n_{\text{critic}}$}
    {
        Sample a batch of $b$ finished images from the trajectory buffer\;
        Sample a batch of $b$ target images from $\mathcal{D}_{\text{target}}$\;
        $w\leftarrow w - \alpha_w \grad_w L_w$\;
    }
    Draw a batch $B$ of $b$ images from the trajectory buffer\;
    Delete images in the batch that are already finished\;
    Refill deleted images in the batch using those from $D_{\text{RAW}}$\;
    Apply one step of operation to the images: $B'=\text{Actor}(B)$\;
    $\theta_1\leftarrow \theta_1 + \alpha_{\theta} \grad_{\theta_1}J(\pi_{\theta})$\;
    $\theta_2\leftarrow \theta_2 + \alpha_{\theta} \grad_{\theta_2}J(\pi_{\theta})$\;
    $v\leftarrow v - \alpha_v \grad_v L_v$\;
    Put new images $B'$ back into the trajectory buffer\;
}
\caption{Training procedure}
\label{alg:train}
\end{algorithm}

\subsection{Quality evaluation via adversarial learning}
To generate results as close to the target dataset as possible, we employ a GAN, which is composed of two parts, namely a generator (i.e., the actor of the previous subsection in our case) and a discriminator. The two parts are optimized in an adversarial manner: the discriminator is trained to tell if the image is from the target dataset or was generated by the generator; the actor aims to ``fool'' the discriminator by generating results as close to the target dataset as possible, so that the discriminator cannot distinguish the difference. The two networks are trained simultaneously, and an ideal equilibrium is achieved when the generated images are close to the targets.

In this work, we use a popular variant of the traditional GAN called the Wasserstein GAN (WGAN)~\cite{arjovsky2017wasserstein}, which uses the Earth Mover's Distance (EMD) to measure the difference between two probability distributions. It has been shown to stabilize GAN training and avoid vanishing gradients. The loss for the discriminator\footnote{The discriminator is referred to as the ``critic'' in~\cite{arjovsky2017wasserstein}. We use the term ``discriminator'' here to distinguish it from the critic in our actor-critic framework.} $D$ is defined as
\begin{equation}
L_w=\E_{s\in \rho^{\policy}}\left[D(s)\right] - \E_{s\in \text{target dataset}}\left[D(s)\right].
\end{equation}
The discriminator $D$ is modeled as a CNN with parameters denoted as $w$. The ``negative loss'' (quality score) for the generator, whose increment serves as a component of the reward in our system, is
\begin{equation}
-L_{\text{actor}}=\E\left[D(s)\right].
\end{equation}
Intuitively, the discriminator aims to give large values for images in the target collection, and small ones to retouched images produced by the generator. On the other hand, the actor (generator) aims to submit an output at a state where the discriminator gives a larger value, meaning that the final image appears more similar to those in the target dataset. Following~\cite{gulrajani2017improved}, we use a gradient penalty instead of weight clipping in the discriminator.

\subsection{Training strategies}
\label{subsec:penalizing}
Both RL algorithms and GANs are known to be hard to train. To address this issue, we utilized the following strategies to stabilize the training process.

\paragraph{Exploitation vs. exploration}
A well-known tradeoff exists between exploitation and exploration, namely whether to devote more attention on improving the current policy or to try a new action in search of potentially greater future reward. This is especially challenging for our two-stage \Changed{decision making}{decision-making} problem, as focusing on one filter may lead to under-exploitation of filter parameter learning for other filters. To avoid such local minima, we penalize $\policy_1$ if its action proposal distribution is too concentrated, i.e., has low entropy. This is done by reducing its reward:
\begin{equation}
R'=R-0.05\left(\log|\mathcal{F}|+\sum_{F\in\mathcal{F}}\policy_1(F)\log \policy_1(F)\right).
\end{equation}

In addition, we found that the agent may use a filter repeatedly during a retouching process, such as by applying two consecutive exposure adjustments rather than combining them into a single step. For a more concise retouching solution, we ``teach'' the agent to avoid actions like this by penalizing filter reuse: if the agent uses a filter twice, the second usage will incur an additional penalty of $-1$. To implement this, the agent needs to know what filters have been applied earlier in the process, so we append this usage information as additional channels in the image (denoted as {\em state planes} in Fig.~\ref{fig:network}). Encouraging the agent to exploit each filter to its maximum potential also leads to greater exploration of different filters.

\paragraph{Out-of-order training}
\changed{}{Images along a single trajectory at consecutive steps can be highly correlated, and such correlation is harmful to both RL and GANs.
To address this, we propose an {\em out-of-order} training scheme rather than sequential training, as detailed in Algorithm~\ref{alg:train}.
Specifically, instead of starting and finishing a small number (e.g., $64$, as a single batch) of trajectories simultaneously,
we maintain a large number of on-the-fly trajectories in a {\em trajectory buffer}. In each training iteration, we sample a batch of images from the trajectory buffer (not necessarily at the same stage), apply one operation step to them, and put the edited images back to the buffer.
The benefits of such a mechanism are two-fold: (1) for RL, it plays partly the role of the {\it experience replay}~\cite{lin1993reinforcement} mechanism, which is observed to ``smooth'' the training data distribution~\cite{mnih2013playing};
(2) for GAN training, this approach acts in a similar manner as the ``history'' buffer in~\cite{shrivastava2017learning}, which also helps to reduce model oscillation.}

\section{Results}

In this section, we present implementation details, a validation, and applications of our system.

\paragraph{Implementation details} TensorFlow~\cite{tensorflow2015-whitepaper}, a deep learning framework which provides automatic differentiation, is used to implement our system. Following the design of filters presented in Section~\ref{sec:filter}, the retouching steps are represented as basic differentiable arithmetic operations. For estimation of retouching actions and parameters, we downsample the high-resolution input image to $64\times 64 px$.
\changed{}{Though the input resolution is not high, it leads to an effective balance between performance and network size. Such resolution provides sufficient information for global retouching parameter estimation; at the same time, the resulting small network size prevents overfitting, leads to fast inference speed, and makes it easy to incorporate the model within applications.}
The estimated actions and parameters are subsequently applied to the full-resolution image at run time.

All networks are optimized using Adam~\cite{kingma2014adam}, with a base learning rate of $1.5\times 10^{-5}$ for the policy networks, $5\times 10^{-5}$ for the discriminator, and $5\times 10^{-4}$ for the value network. During training, these learning rates are exponentially decayed to $10^{-3}$ of the original values. \changed{}{We use a larger learning rate and more iterations for the discriminator \Changed{}{than for the generator}, to make sure it ``saturates'' and keeps generating a relatively tight lower bound of the EMD.} Training takes less than $3$ hours for all the experiments.

\paragraph{Efficiency and model size}
Thanks to the resolution-independent filter design, the computation of our method is fast: an unoptimized version takes $30ms$ for inference on an NVIDIA TITAN X (Maxwell) GPU. The model size is small ($<30MB$) and therefore can be conveniently shipped with a mobile application or digital camera. This opens up the possibility of providing users with automatically retouched images in the camera viewfinder in real time.

\paragraph{Datasets} We utilize two sources of training data:

\begin{itemize}
\item{The MIT-Adobe FiveK Dataset.} Bychkovsky \etal~\shortcite{bychkovsky2011learning} compiled a photo dataset consisting of $5,000$ RAW images and retouched versions of each by five experts. In this work, we randomly separate the dataset into three parts: \textbf{(part 1)} $2,000$ input RAW images, \textbf{(part 2)} $2,000$ retouched images by retoucher C, and \textbf{(part 3)} $1,000$ input RAW images for testing. The three parts have no intersection with each other.

\item{The 500px Dataset.}
We crawled professionally retouched photos from two artists on \url{500px.com}. The two sets of data have relatively consistent styles, and are comprised of $369$ and $397$ photos each.
\end{itemize}

\paragraph{Error Metrics}
The novel learning framework enables our system to take advantage of unpaired training data, which requires error metrics different from previous work.

It is shown in \cite{hwang2012context} and that the $l_2$ loss may not accurately reflect visual quality. This problem is especially apparent when a dataset exhibits multi-modal appearance or style, such as black-and-white apples retouched into red (e.g., $RGB=(1,0,0)$) or green ($RGB=(0, 1, 0)$) with a $50\%$ probability for each. As pointed out in \cite{pathak2016context, zhang2016colorful, isola2016image}, use of simple loss functions like $l_1$ or $l_2$ can lead to ``blurry'' results for image generation. For the \Changed{apples}{apple} example, a CNN with an $l_2$ loss will end up generating yellow ($RGB=(0.5, 0.5, 0)$) apples, which minimizes the loss but may produce styles that do not even exist in the dataset. Multi-modality naturally exists in retouching, since the same artist may retouch an image in different ways. The inconsistent nature of retouching is exhibited in the MIT-Adobe FiveK dataset and was also observed in \cite{yan2016automatic}.

Therefore, even if input-output pairs do exist for some tasks, the $l_2$ loss may not be a suitable learning metric. However, how to automatically evaluate the perceptual quality of style learning remains an open problem. Though such metrics are difficult to design and are sometimes unreliable, for development and debugging purposes, it would still be good to have an automatic way to roughly measure how well the model fits the target data. Toward this end, we evaluate the similarity of generated images to target images based on their {\em distributions} of image properties. In~\cite{isola2016image}, the distances of L, a, b distributions are measured using the intersections of their histograms in the output and target datasets. In our work, we use \textbf{luminance}, \textbf{contrast}, \textbf{saturation} as three descriptive features of image styles, and measure the distance of their distributions in the output and target images using histogram intersections. A detailed explanation of this metric is given in the supplemental document~\cite{hu2018b}.

In addition to histogram intersections, we employ user studies via Amazon Mechanical Turk (AMT) for perceptual evaluation of this work. For each group of outputs from a given method, we randomly choose $100$ images and ask users to rate \Changed{the image}{them}. The user is presented with one output image (with target style image thumbnails, if necessary) at a time and is prompted to give a score from $1$ (worst) to $5$ (best) to each image, based on image quality and style. $5$ ratings are collected for each image, resulting in $500$ ratings for each group of outputs. Please refer to the supplemental document~\cite{hu2018b} for more details about our AMT experiments.

\subsection{End-to-end Post-Processing and Style Learning}
\newcommand{\gallerycell}[2]{{\setlength{\fboxsep}{0pt}\color{white}\setlength{\fboxrule}{1pt}\fbox{\includegraphics[width=0.092\textwidth]{images/gallery1/#2_#1}}\color{black}}}
\newcommand{\galleryimage}[1]{

\gallerycell{#1}{expert} &
\gallerycell{#1}{CycleG} &
\gallerycell{#1}{pix2pi} &
\gallerycell{#1}{human} &
\gallerycell{#1}{dump-k}\\
}
\begin{figure}
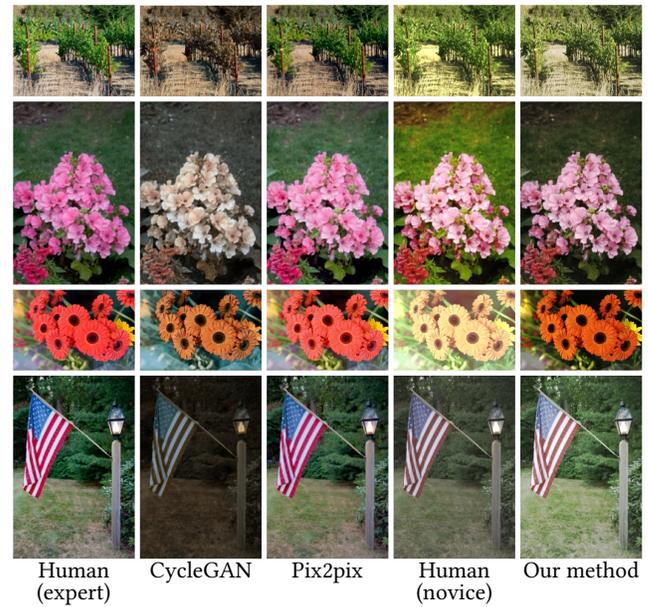

\begin{center}
\scalebox{0.98}{
\begin{zerotabular}{ccccc}
\galleryimage{0153}
\galleryimage{2145}
\galleryimage{3276}
\galleryimage{4123}
Human & CycleGAN & Pix2pix & Human & Our method \\
(expert) & & & (novice) & \\
\end{zerotabular}
}
\end{center}
\caption{Results of \Changed{a} human expert, CycleGAN, Pix2pix, human \Changed{novice}{novices}, and our method on the MIT-Adobe FiveK Dataset.}
\label{fig:gallery}
\end{figure}

\begin{figure}
  \centering
  \resizebox{1.0\linewidth}{!}{
  \begin{tikzpicture}[scale=0.67]
    \node[inner sep=0pt] (input) at (-4,-3) {\includegraphics[width=0.14\textwidth]{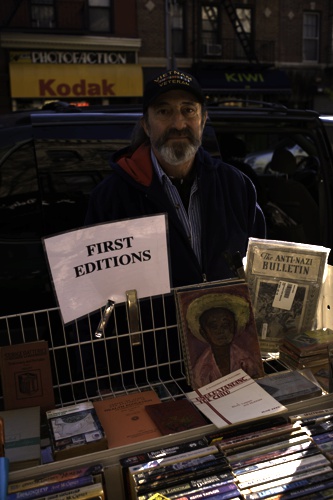}};
    \node (inputtag) at ([yshift=-1.4em] input.south) {\textbf{Input}};
    \node[inner sep=0pt] (cyc) at (0,-3) {\includegraphics[width=0.14\textwidth]{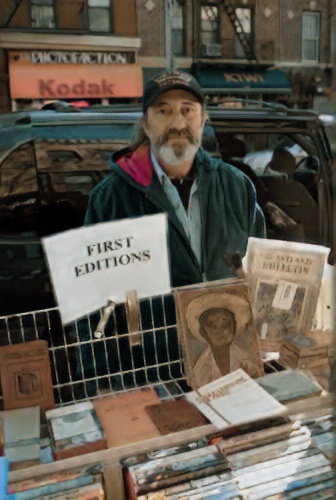}};
    \node[inner sep=0pt, draw=red, very thick] (cycmax) at (0,-9.2) {\includegraphics[width=0.14\textwidth]{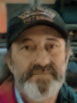}};
    \node[inner sep=0pt] (p2p) at (4,-3) {\includegraphics[width=0.14\textwidth]{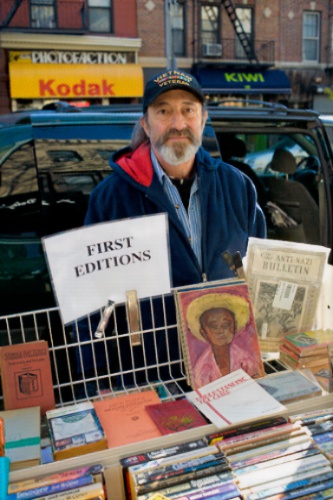}};
    \node[inner sep=0pt, draw=red, very thick] (p2pmax) at (4,-9.2) {\includegraphics[width=0.14\textwidth]{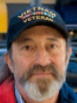}};
    \node[inner sep=0pt] (ours) at (8,-3) {\includegraphics[width=0.14\textwidth]{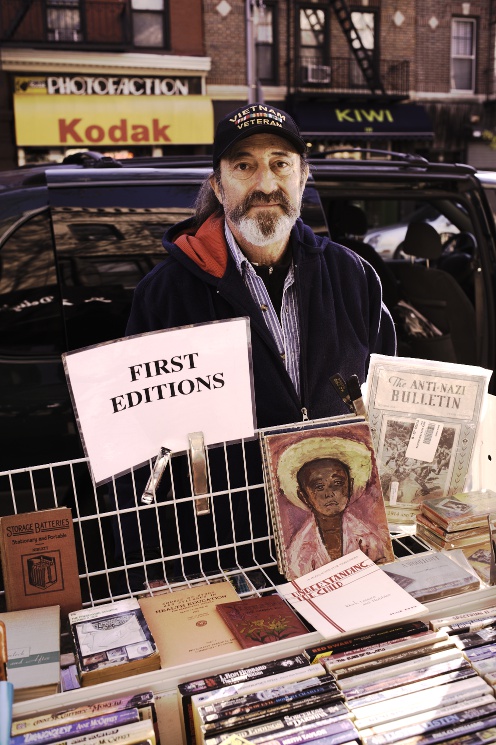}};
    \node[inner sep=0pt, draw=red, very thick] (oursmax) at (8,-9.2) {\includegraphics[width=0.137\textwidth]{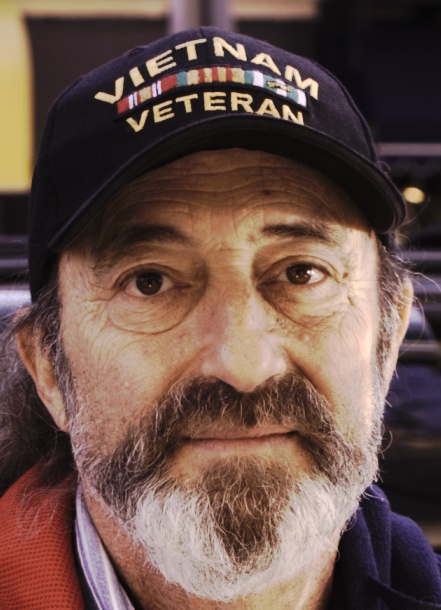}};
    \node (cyctag) at ([yshift=-1.4em] cyc.south) {\textbf{CycleGAN}};
    \node (p2ptag) at ([yshift=-1.4em] p2p.south) {\textbf{Pix2pix}};
    \node (ourstag) at ([yshift=-1.4em] ours.south) {\textbf{Ours}};
    \node (p2pmaxtag) at ([yshift=-1.4em] p2pmax.south) {\textbf{Zoom-in views}};
  \end{tikzpicture}
  }
  \caption{Comparison to CycleGAN and Pix2pix results. While they can produce good tone and color compared to the input, our method additionally has no distortion and no limit on image resolution.}
  \label{fig:cyclegan}
\end{figure}

\label{sec:cyclegan}
Most previous methods for automatic photo post-processing are based on supervised learning which requires paired data~\cite{dale2009image, bychkovsky2011learning, hwang2012context, yan2016automatic, gharbi2017deep}. It is only recently that a series of works~\cite{liu2016coupled, shrivastava2017learning, zhu2017unpaired, kim2017learning} based on GANs have made possible the utilization of unpaired data. We compare our results with those of CycleGAN, another deep learning approach for image generation using only unpaired data. Note that in contrast to CycleGAN, our method has no limitation on resolution, since the filters are resolution-independent and filter operations estimated from low-res images can be identically applied to high-res inputs.

We conducted three sets of experiments using RAW images from \textbf{part 1} of the MIT-Adobe FiveK dataset as input. For the target datasets we use images from expert C \Changed{}{\textbf{(part 2)}} in the MIT-Adobe FiveK dataset and the two artists from 500px.com, respectively.
Though Pix2pix~\cite{isola2016image} needs {\em paired} data to work, we still include its performance on a test using paired data from \textbf{part 1} of the MIT-Adobe FiveK dataset, \Changed{}{retouched by expert C}.

\newcommand{\addHist}{1}{}

\begin{table}[b]
\caption{Quantitative results on general post-processing (using expert C in MIT-Adobe FiveK as training target dataset).}
\label{tab:benchmark_A}
\scalebox{0.96}{
\hspace{-2.9mm}
\begin{minipage}{\columnwidth}
\begin{tabular}{l|ccc|c}
  \toprule
  \textbf{Approach}& \multicolumn{3}{c|}{\textbf{Histogram Intersection}} & \textbf{AMT} \\
   & Luminance & Contrast & Saturation &  \textbf{User Rating} \\ \hline
  Ours & $71.3\%$ & $83.7\%$& $69.7\%$& $3.43$ \\ \hline
  CycleGAN & $61.4\%$&$71.1\%$ &$82.6\%$ & $2.47$\\ \hline
  Pix2pix &$92.4\%$ &$83.3\%$ &$86.5\%$ & $3.37$ \\ \hline \hline
  Human & - & - & - & $3.30$\\ \hline
  Expert C & $100\%$& $100\%$&$100\%$ & $3.66$\\
  \bottomrule
\end{tabular}
\begin{overpic}[width=1.0\linewidth]{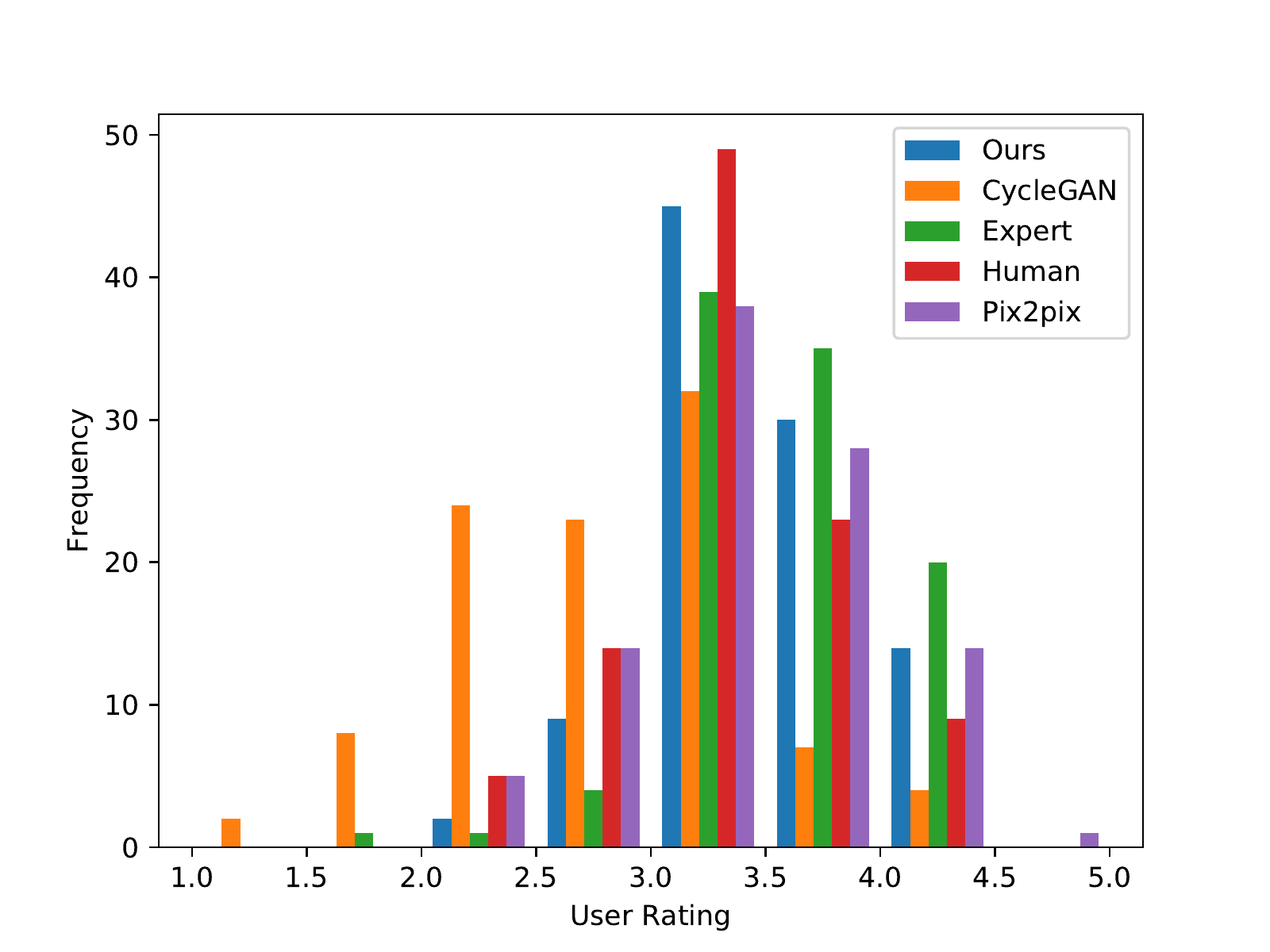}
 \put (28,68) {\textbf{Image Rating Distributions}}
\end{overpic}
\end{minipage}
}
\end{table}

\begin{figure*}
   \centering
   \resizebox{0.97\linewidth}{!}{
   \begin{zerotabular}{cccc}
   \GGO{fk_C}{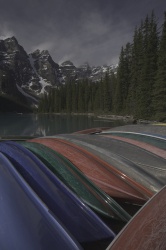} &
   \GGO{fk_C}{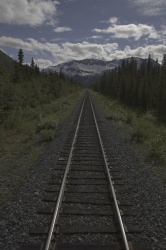} \\
   \GGO{fk_C}{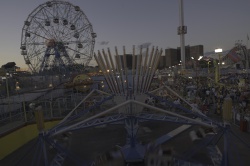} &
   \GGO{fk_C}{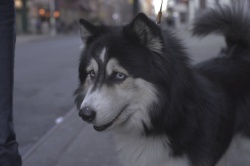} \\
   \GGO{fk_C}{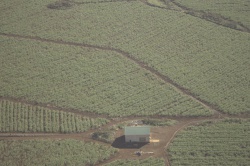} &
   \GGO{fk_C}{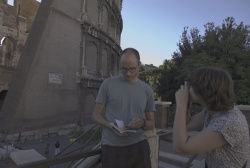} \\
   \GGO{fk_C}{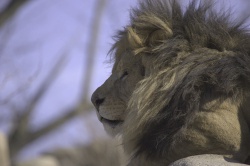} &
   \GGO{fk_C}{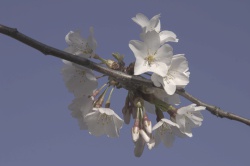} \\
   Input (tone-mapped) & Retouched & Input (tone-mapped) & Retouched
   \end{zerotabular}
   }
  \caption{Photos retouched by our system, trained on expert C from the MIT-Adobe FiveK dataset.}
  \label{fig:final1}
\end{figure*}

For the first experiment with images from expert C as target images, visual results are shown in Figure~\ref{fig:gallery}, ~\ref{fig:cyclegan} and ~\ref{fig:final1}, and quantitative results are listed in Table~\ref{tab:benchmark_A}. It can be seen that Pix2pix and CycleGAN generate vivid color but lead to edge distortions and degraded image quality, making them unsuitable for high-quality post-processing tasks.
Using the publicly available implementation of CycleGAN from the authors~\cite{zhu2017unpaired}, training takes $30$ hours for generating images of resolution $500 \times 333$px\footnote{We used {\em fineSize}=$128$ in the authors' implementation (\url{https://github.com/junyanz/CycleGAN}).}.

For the style learning experiments with the {\it 500px} artists, quantitative results are shown in Table~\ref{tab:benchmark_fhp_A} and Table~\ref{tab:benchmark_fhp_B}, and visual results are displayed in Figure~\ref{fig:style-learning}. No comparison results can be generated for Pix2pix, since no paired training data is generally available for images downloaded from the web.

\begin{table}
\caption{Quantitative results on style learning (using artist A in 500px as training target dataset).}
\label{tab:benchmark_fhp_A}
\scalebox{0.96}{
\hspace{-2.9mm}
\begin{minipage}{\columnwidth}
\begin{tabular}{l|ccc|c}
  \toprule
  \textbf{Approach}& \multicolumn{3}{c|}{\textbf{Histogram Intersection}} & \textbf{AMT} \\
   & Luminance & Contrast &  Saturation& \textbf{User Rating} \\ \hline
  Ours & $82.4\%$ & $80.0\%$& $71.5\%$& $3.39$ \\ \hline 
  CycleGAN & $63.6\%$&$45.2\%$ &$71.8\%$ & $2.69$\\ \hline \hline
  500px artist A & $100\%$& $100\%$&$100\%$ & $3.72$\\
  \bottomrule
\end{tabular}
\begin{overpic}[width=1.0\linewidth]{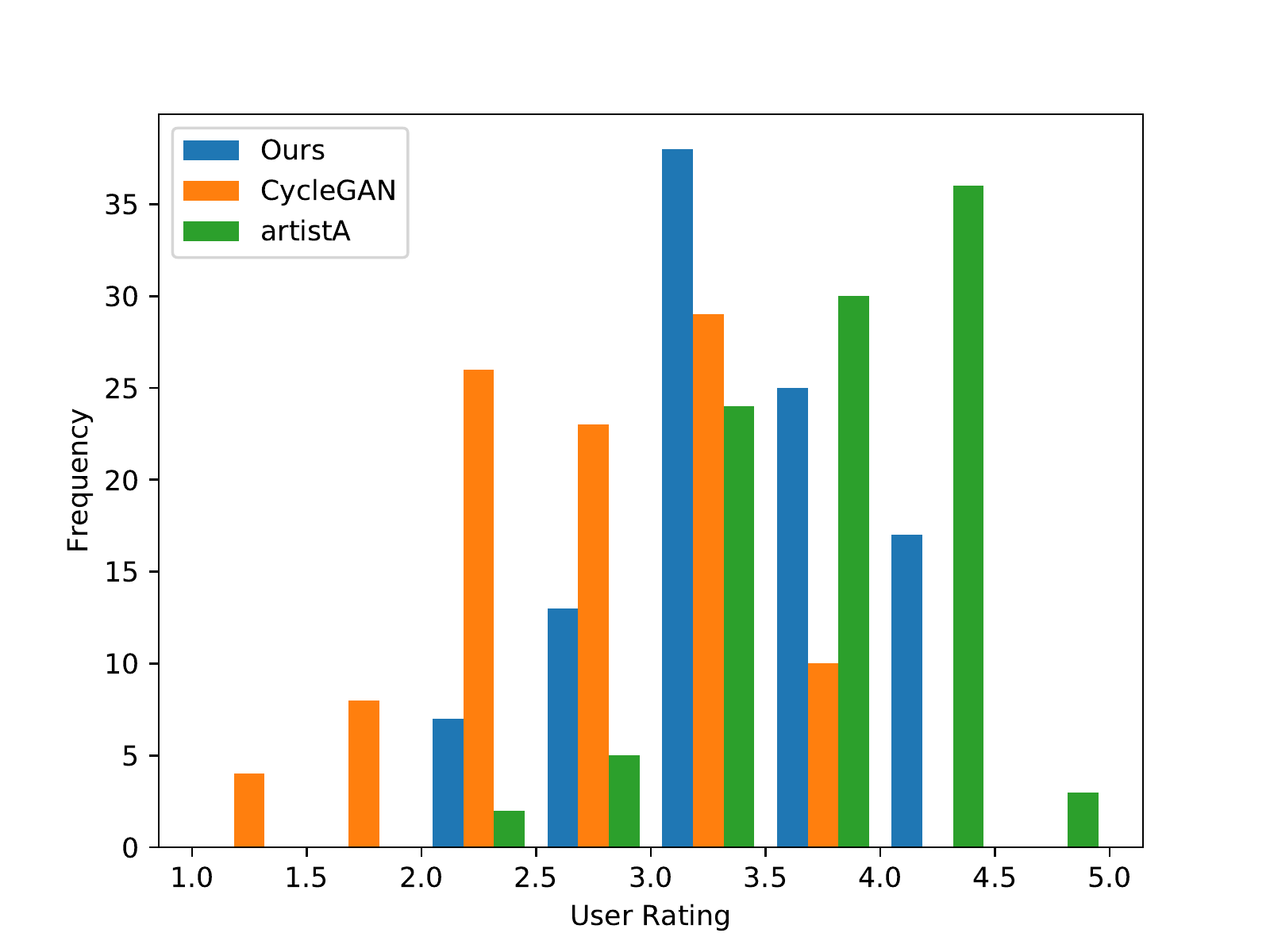}
 \put (28,68) {\textbf{Image Rating Distributions}}
\end{overpic}
\end{minipage}
}
\end{table}

\begin{table}
\caption{Quantitative results on style learning (using artist B in 500px as training target dataset).}
\label{tab:benchmark_fhp_B}
\scalebox{0.96}{
\hspace{-2.9mm}
\begin{minipage}{\columnwidth}
\begin{tabular}{l|ccc|c}
  \toprule
  \textbf{Approach}& \multicolumn{3}{c|}{\textbf{Histogram Intersection}} & \textbf{AMT} \\
   & Luminance & Contrast & Saturation & \textbf{User Rating} \\ \hline
  Ours & $85.2\%$ & $91.7\%$& $83.5\%$& $3.22$ \\ \hline 
  CycleGAN & $60.1\%$&$79.4\%$ &$83.4\%$ & $2.86$\\ \hline \hline
  500px artist B & $100\%$& $100\%$&$100\%$ & $3.40$\\
  \bottomrule
\end{tabular}
\begin{overpic}[width=1.0\linewidth]{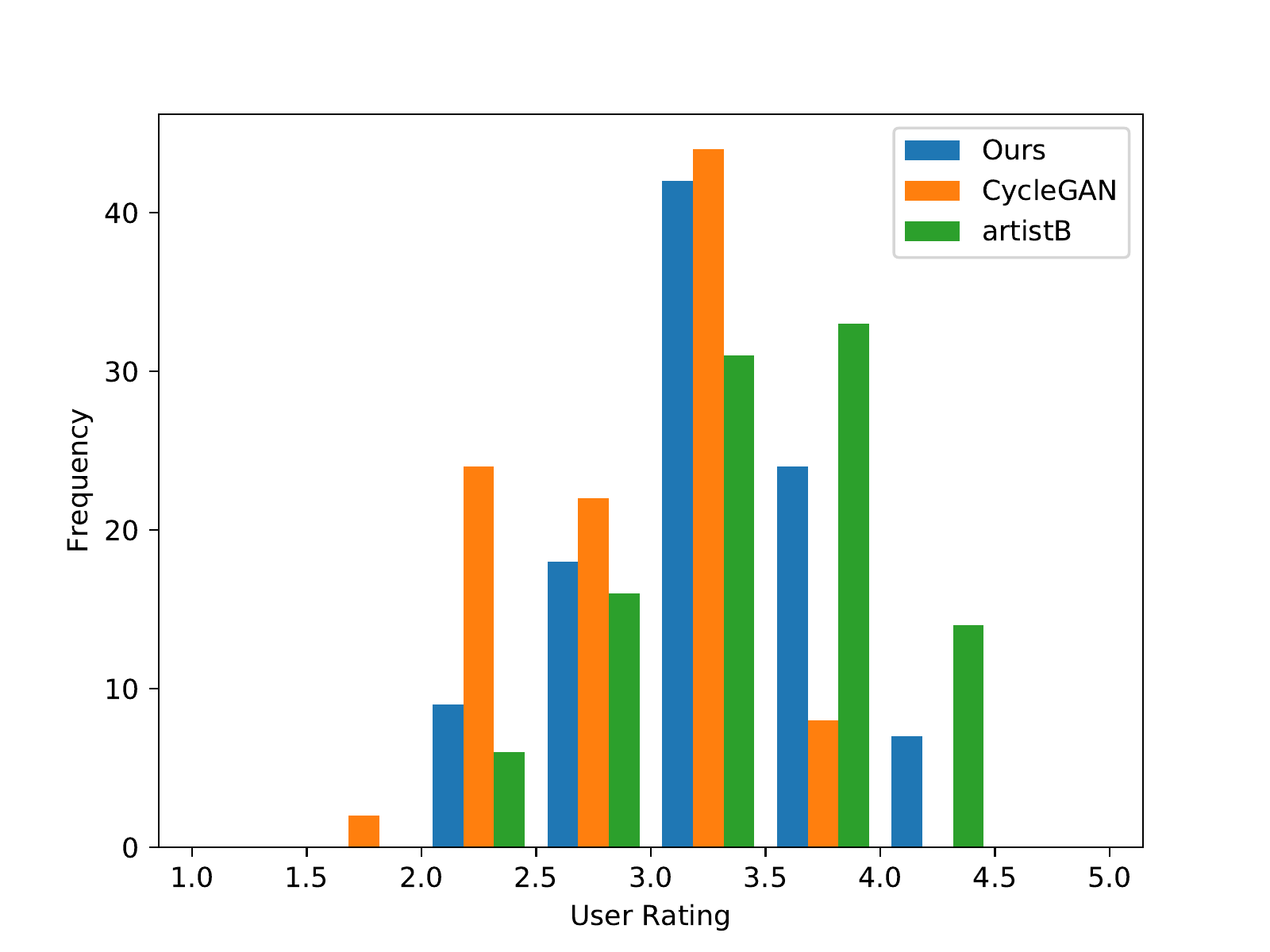}
 \put (28,68) {\textbf{Image Rating Distributions}}
\end{overpic}
\end{minipage}
}
\end{table}

\setlength{\tabcolsep}{2pt}

\begin{figure*}
   \centering
  \resizebox{0.97\linewidth}{!}{
   \begin{zerotabular}{cccccc}
   \GGOH{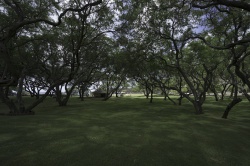} &
   \GGOH{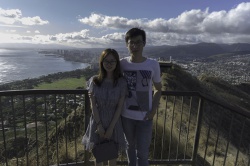} &
   \GGOH{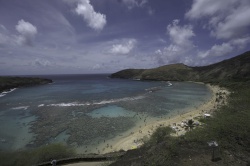}
   \\
   \GGOH{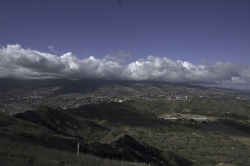} &
   \GGOH{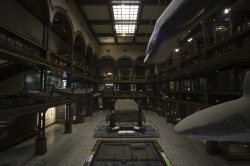} &
   \GGOH{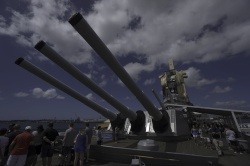}
   \\
   Input (tone-mapped) & Retouched & Input (tone-mapped) & Retouched & Input (tone-mapped) & Retouched
   \end{zerotabular}}
  \caption{
  We apply our trained model (artist B) to another RAW photo dataset, to validate its cross-dataset generalization performance.}
  \label{fig:generalization}
\end{figure*}

\newcommand{\styleWidth}{0.592\textwidth}
\newcommand{\styleHeight}{0.108\textwidth}
\newcommand{\styleYa}{-5}
\newcommand{\styleYb}{-7.0}
\newcommand{\styleYc}{-18.5}
\newcommand{\styleYd}{-20.5}

\newcommand{\styleImgA}{0476}
\newcommand{\styleImgB}{1370}
\newcommand{\styleImgC}{1298}
\newcommand{\styleImgD}{1318}

\LARGE

\begin{figure}
  \resizebox{1.00\linewidth}{!}{
   \begin{tikzpicture}[scale=1.0]
        
        \draw[thick] (-0.4,-8.33) -- (10.84,-8.33);
        \draw[thick] (-0.4,-8.4) -- (10.84,-8.4);

        \node[inner sep=0pt] (style_A) at (5.5759,0.28) {\includegraphics[trim={0 5cm 0 0},clip,width=\styleWidth]{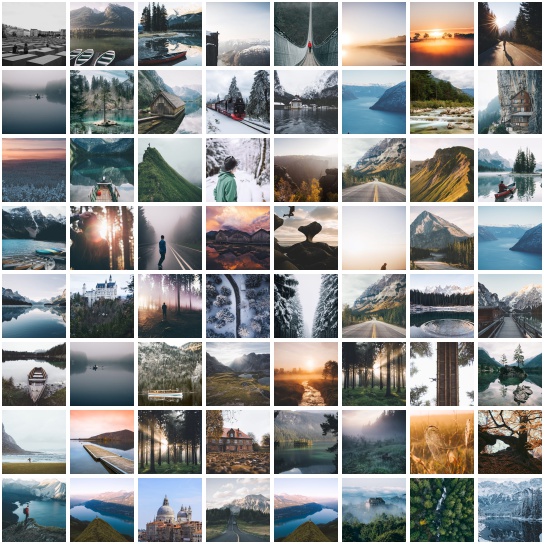}};
        \node[inner sep=0pt] (style_B) at (5.5759,-13.22) {\includegraphics[trim={0 5cm 0 0},clip,width=\styleWidth]{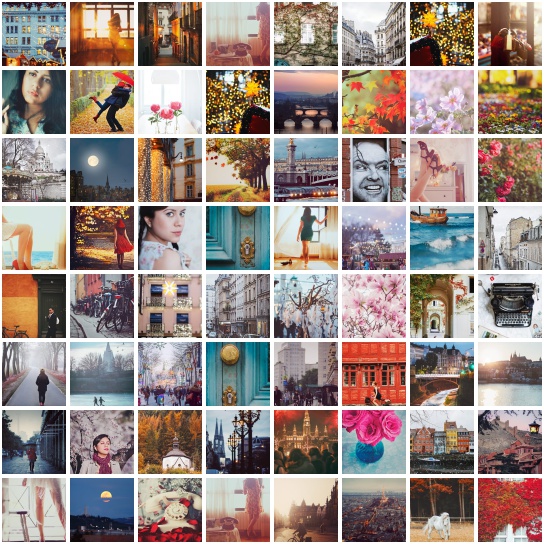}};

        \node (style_A_input_label) at ([xshift=-0.8em]style_A.west) {\rotatebox{90}{Target photos}};
        \node (style_B_input_label) at ([xshift=-0.8em]style_B.west) {\rotatebox{90}{Target photos}};

        \node (style_A_label) at ([yshift=1.0em]style_A.north)
{\rotatebox{0}{\textbf{500px Artist A}}};
        \node (style_B_label) at ([yshift=1.0em]style_B.north)
{\rotatebox{0}{\textbf{500px Artist B}}};

        \node[inner sep=0pt] (style_A_ours1) at (2,\styleYa) {\includegraphics[height=\styleHeight]{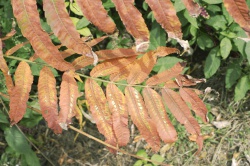}};
        \node[inner sep=0pt] (style_A_ours2) at (5,\styleYa) {\includegraphics[height=\styleHeight]{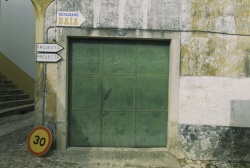}};
        \node[inner sep=0pt] (style_A_ours3) at (8,\styleYa) {\includegraphics[height=\styleHeight]{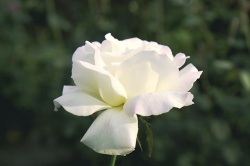}};
        \node[inner sep=0pt] (style_A_ours4) at (10.2,\styleYa) {\includegraphics[height=\styleHeight]{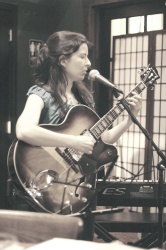}};

        \node[inner sep=0pt] (style_A_cyc1) at (2,\styleYb) {\includegraphics[height=\styleHeight]{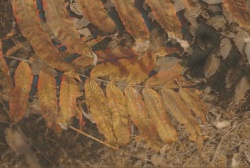}};
        \node[inner sep=0pt] (style_A_cyc2) at (5,\styleYb) {\includegraphics[height=\styleHeight]{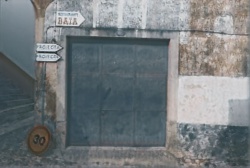}};
        \node[inner sep=0pt] (style_A_cyc3) at (8,\styleYb) {\includegraphics[height=\styleHeight]{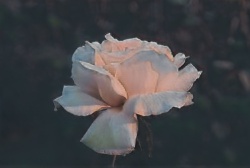}};
        \node[inner sep=0pt] (style_A_cyc4) at (10.2,\styleYb) {\includegraphics[height=\styleHeight]{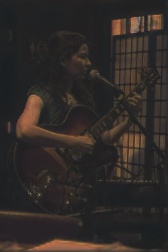}};

        \node (style_A_ours_label) at ([xshift=-0.5em]style_A_ours1.west) {\rotatebox{90}{Ours}};
        \node (style_A_cyc_label) at ([xshift=-0.5em]style_A_cyc1.west) {\rotatebox{90}{CycleGAN}};
        \node (style_A_outputs_label) at ([xshift=-1.6em, yshift=-3em]style_A_ours1.west) {\rotatebox{90}{Outputs}};

        \node[inner sep=0pt] (style_B_ours1) at (2,\styleYc) {\includegraphics[height=\styleHeight]{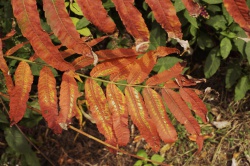}};
        \node[inner sep=0pt] (style_B_ours2) at (5,\styleYc) {\includegraphics[height=\styleHeight]{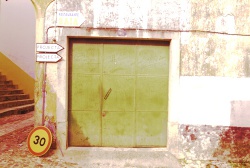}};
        \node[inner sep=0pt] (style_B_ours3) at (8,\styleYc) {\includegraphics[height=\styleHeight]{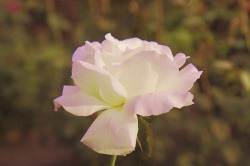}};
        \node[inner sep=0pt] (style_B_ours4) at (10.2,\styleYc) {\includegraphics[height=\styleHeight]{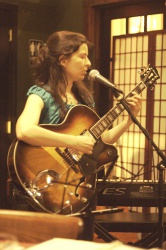}};

        \node[inner sep=0pt] (style_B_cyc1) at (2,\styleYd) {\includegraphics[height=\styleHeight]{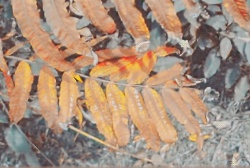}};
        \node[inner sep=0pt] (style_B_cyc2) at (5,\styleYd) {\includegraphics[height=\styleHeight]{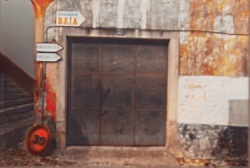}};
        \node[inner sep=0pt] (style_B_cyc3) at (8,\styleYd) {\includegraphics[height=\styleHeight]{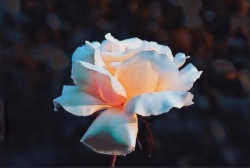}};
        \node[inner sep=0pt] (style_B_cyc4) at (10.2,\styleYd) {\includegraphics[height=\styleHeight]{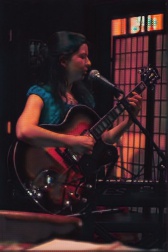}};

        \node (style_B_ours_label) at ([xshift=-0.5em]style_B_ours1.west) {\rotatebox{90}{Ours}};
        \node (style_B_cyc_label) at ([xshift=-0.5em]style_B_cyc1.west) {\rotatebox{90}{CycleGAN}};
        \node (style_B_outputs_label) at ([xshift=-1.6em, yshift=-3em]style_B_ours1.west) {\rotatebox{90}{Outputs}};

    \end{tikzpicture}
    }
  \caption{Learning the styles of two artists from 500px.com, using our system and CycleGAN.}
  \label{fig:style-learning}
\end{figure}

\normalsize

\paragraph{Generalization} \changed{}{Since the number of training images is small, it is worth investigating the generalization ability of our model.
To this end, we apply the network to another set of RAW photos. Promising results are obtained, as shown in Figure~\ref{fig:generalization}.
}

\begin{figure}
  \centering
  \scalebox{1.0}{
    \begin{zerotabular}{ccc}
    \GGG{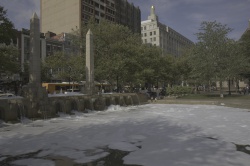} \\
    \GGG{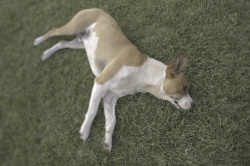} \\
    \GGG{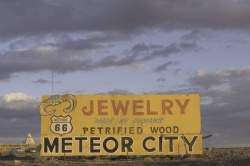} \\
    \GGG{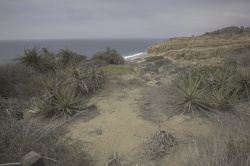} \\
    \GGG{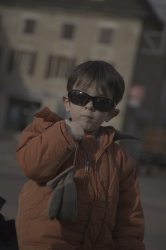} \\
    \GGG{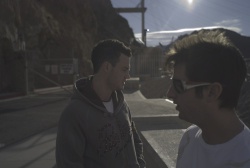} \\
    \GGG{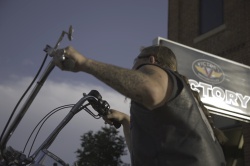} \\
    \GGG{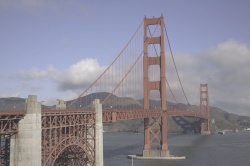} \\
    \GGG{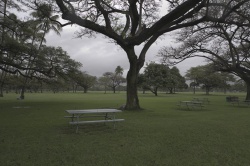} \\ \\
    Input (tone-mapped) & Stylized (artist A) & Stylized (artist B) \\
    \end{zerotabular}
  }
  \caption{Photos retouched by our system, trained on two artists from {\em 500px}.}
  \label{fig:final12}
\end{figure}

\setlength{\tabcolsep}{4pt}
\renewcommand{\folder}{1730A}
\begin{figure}
    \centering
    \resizebox{0.86\linewidth}{!}{
    \begin{tikzpicture}[scale=0.95]
        \node[inner sep=0pt] (image1) at (0,0) {\includegraphics[width=.19\textwidth]{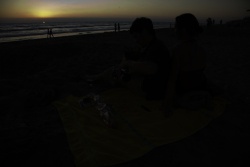}};
        \node[inner sep=0pt] (image2) at (0,-3) {\includegraphics[width=.19\textwidth]{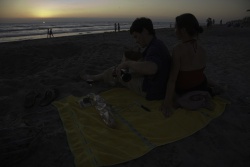}};
        \node[inner sep=0pt] (image3) at (0,-6) {\includegraphics[width=.19\textwidth]{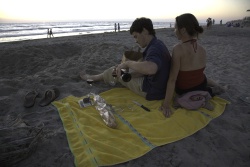}};
        \node[inner sep=0pt] (image4) at (0,-9) {\includegraphics[width=.19\textwidth]{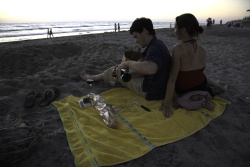}};
        \node[inner sep=0pt] (image5) at (0, -12) {\includegraphics[width=.19\textwidth]{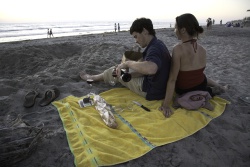}};
        \node[inner sep=0pt] (image6) at (1.75, -16.5) {\includegraphics[width=.372\textwidth]{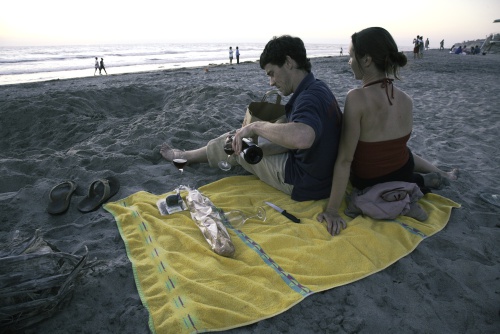}};

        \foreach \from/\to in {image1/agent1, image2/agent2, image3/agent3, image4/agent4, image5/agent5}
            \draw[->,very thick] (\from.east) -- (\to.west);
        \foreach \from/\to in {agent1/image2, agent2/image3, agent3/image4, agent4/image5, agent5/image6}
            \draw[->,very thick] (\from.south) -- ++(0,-0.1) -- ++(0,-0.2) -| (\to.north);
    \end{tikzpicture}
  }
  \caption{Example of a learned retouching operation sequence from artist A (500px).}
  \label{fig:trajFhpA}
\end{figure}

\renewcommand{\folder}{1730B}
\begin{figure}
    \centering
    \resizebox{0.86\linewidth}{!}{
    \begin{tikzpicture}[scale=0.95]
        \node[inner sep=0pt] (image1) at (0,0) {\includegraphics[width=.19\textwidth]{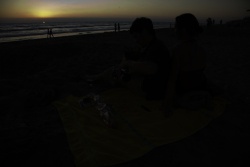}};
        \node[inner sep=0pt] (image2) at (0,-3) {\includegraphics[width=.19\textwidth]{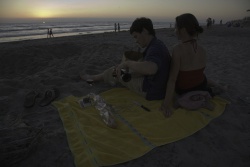}};
        \node[inner sep=0pt] (image3) at (0,-6) {\includegraphics[width=.19\textwidth]{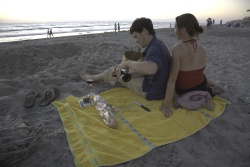}};
        \node[inner sep=0pt] (image4) at (0,-9) {\includegraphics[width=.19\textwidth]{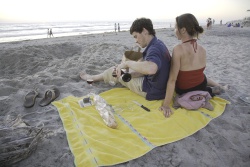}};
        \node[inner sep=0pt] (image5) at (0, -12) {\includegraphics[width=.19\textwidth]{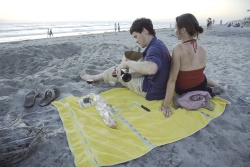}};
        \node[inner sep=0pt] (image6) at (1.75, -16.5) {\includegraphics[width=.372\textwidth]{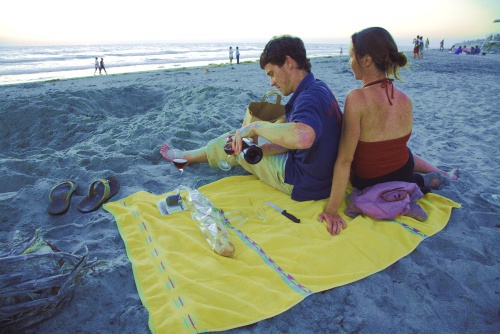}};

        \foreach \from/\to in {image1/agent1, image2/agent2, image3/agent3, image4/agent4, image5/agent5}
            \draw[->,very thick] (\from.east) -- (\to.west);
        \foreach \from/\to in {agent1/image2, agent2/image3, agent3/image4, agent4/image5, agent5/image6}
            \draw[->,very thick] (\from.south) -- ++(0,-0.1) -- ++(0,-0.2) -| (\to.north);
    \end{tikzpicture}
  }
  \caption{Example of a learned retouching operation sequence from artist B (500px). Note that different from artist A,
  photos from artist B are more saturated, which is reflected in this learned operation sequence.}
  \label{fig:trajFhpB}
\end{figure}

\paragraph{Discussion} It can be seen that our method outperforms strong baselines in the user study, with higher user ratings than Pix2pix (which relies on much stronger supervision from paired training data), likely due to the fact that our images have no blurry artifacts. CycleGAN, the unpaired version of Pix2pix, does not perform as well, likely due to much weaker supervision from only unpaired data. It is worth noting that during the user study, the image resolution we used was around $500\times 333$px, so that the resolution problem of CycleGAN and Pix2pix may not be very pronounced. However, at higher output resolutions, it becomes clear that our method generates images of much higher quality, as shown in Figure~\ref{fig:cyclegan}. The deconvolution structure of CycleGAN and Pix2pix enable them to generate structural transformations of images, e.g. painting stripes on horses to generate zebras. However, on our task, such a capability can bring distortion artifacts. The {\bf Contrast} histogram intersection score for CycleGAN on the artist A experiment (Table~\ref{tab:benchmark_fhp_A}) is lower than the other metrics. We hypothesize the reason to be that its small receptive field (1/3 of the whole image width) does not adequately capture low-frequency image variations, which is a feature of this artist. A larger receptive field or downsampled image could be used for CycleGAN, but this would require more training data and would produce even lower-resolution outputs.

In conclusion, the results of our system on the retouching problem are very promising. We note though that Pix2pix and CycleGAN can produce extraordinary results on image translation with structural transformations, while our system is tailored for photo post-processing and is not capable of such structural transformations.

\subsection{Reverse Engineering Black-box Filters}
Our work is not the first attempt to mimic the effects of black-box filters.
Previous methods~\cite{yan2016automatic, gharbi2017deep} have shown excellent results in doing so for Instagram/Photoshop filters.
However, these learned filters do not reveal how the original filter works, i.e. we are only getting another \Changed{black-box}{black box} out of an existing one.

\begin{figure}
    \centering
    \resizebox{0.86\linewidth}{!}{
    \begin{tikzpicture}[scale=0.95]
        \node[inner sep=0pt] (image1) at (0,0) {\includegraphics[width=.19\textwidth]{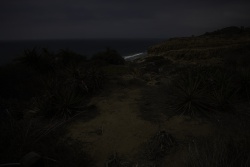}};
        \node[inner sep=0pt] (image2) at (0,-3) {\includegraphics[width=.19\textwidth]{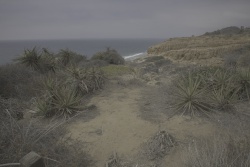}};
        \node[inner sep=0pt] (image3) at (0,-6) {\includegraphics[width=.19\textwidth]{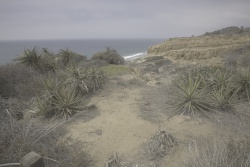}};
        \node[inner sep=0pt] (image4) at (0,-9) {\includegraphics[width=.19\textwidth]{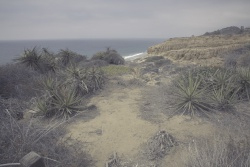}};
        \node[inner sep=0pt] (image5) at (0, -12) {\includegraphics[width=.19\textwidth]{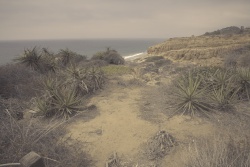}};
        \node[inner sep=0pt] (image6) at (1.75, -16.5) {\includegraphics[width=.372\textwidth]{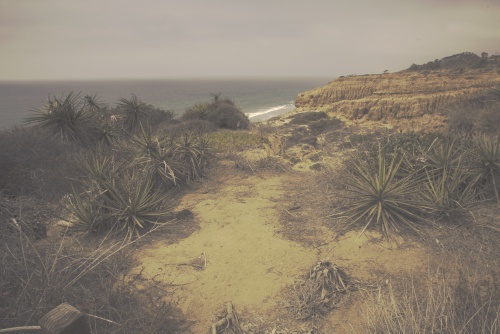}};

        \foreach \from/\to in {image1/agent1, image2/agent2, image3/agent3, image4/agent4, image5/agent5}
            \draw[->,very thick] (\from.east) -- (\to.west);
        \foreach \from/\to in {agent1/image2, agent2/image3, agent3/image4, agent4/image5, agent5/image6}
            \draw[->,very thick] (\from.south) -- ++(0,-0.1) -- ++(0,-0.2) -| (\to.north);
    \end{tikzpicture}
    }
  \caption{Example of a learned operation sequence on the ``Nashville'' filter from Instagram.}
  \label{fig:trajNash}
\end{figure}

Our method not only generates visually pleasing results, but also reveals how this process is done step by step, as shown in Figure~\ref{fig:teaser} and~\ref{fig:trajectory} (on expert C from the MIT-Adobe FiveK dataset), Figure~\ref{fig:trajFhpA} (on artist A from 500px), Figure~\ref{fig:trajFhpB} (on artist B from 500px) and Figure~\ref{fig:trajNash} (on the black-box filter ``Nashville" from Instagram). This is the first time such understandable results can be obtained\changed{}{in a deep learning-based image processing system}, to the best of our knowledge.

\setlength{\tabcolsep}{4pt}
\renewcommand{\folder}{0801}

\begin{figure}
   \centering
   \scalebox{1.0}{
   \begin{zerotabular}{ccccc}
   \multicolumn{5}{c}{\textbf{\includegraphics[width=.45\textwidth]{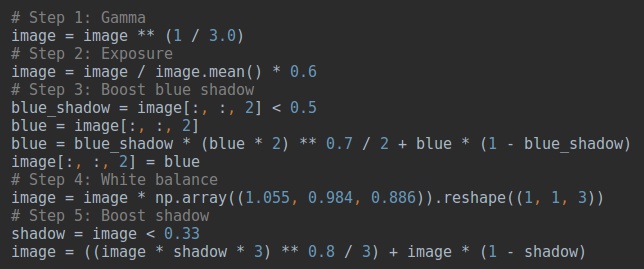}}}\\
   \multicolumn{5}{c}{{Code based on the learned trajectory}}\\
   \includegraphics[width=.09\textwidth]{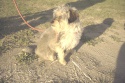} &
   \includegraphics[width=.09\textwidth]{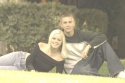} &
   \includegraphics[width=.09\textwidth]{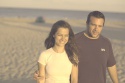} &
   \includegraphics[width=.09\textwidth]{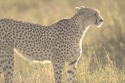} &
   \includegraphics[width=.09\textwidth]{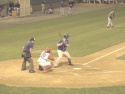}\\
   \multicolumn{5}{c}{{Images generated by the code}}\\
   \includegraphics[width=.09\textwidth]{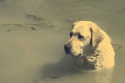} &
   \includegraphics[width=.09\textwidth]{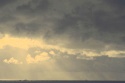} &
   \includegraphics[width=.09\textwidth]{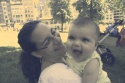} &
   \includegraphics[width=.09\textwidth]{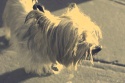} &
   \includegraphics[width=.09\textwidth]{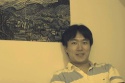}\\
   \multicolumn{5}{c}{{Images generated by the black-box filter}}\\
   \end{zerotabular}
   }
  \caption{With the operation sequence estimated by our system, we can write code that mimics certain black-box filters.}
  \label{fig:code}
\end{figure}

Interestingly, with the help of our system, we can even write explicit code for a black-box filter based on the estimated operation sequence, as illustrated in Figure~\ref{fig:trajNash} and \ref{fig:code}. We believe this capability can greatly help advanced users to gain insight into the artistic styles of particular photographers.


For a given target dataset, the variation among the learned operation sequences reveals how consistent the dataset's image style is.
We find that for the ``Nashville'' filter, the operation sequences are basically the same \changed{}{as in Figure~\ref{fig:trajNash} for all input images}, while for human artists the sequences vary more. This observation matches the previous discussions regarding the error metric and the multi-modal nature of human retouching.

\subsection{Comparison with human users}
Unlike image classification, retouching is a challenging task for most ordinary people. While they can judge how good-looking an image is, it is often challenging for them to generate a nicely retouched photo. Though experts prefer manually retouching photos for maximum control, one of the main goals of our system is to help ordinary users to obtain better photos automatically. Therefore, we examine how normal users perform at this task, and how our system compares to them.

\Changed{We developed a graphical user interface to measure human performance on this task.}{To this end, a graphical user interface (GUI) is developed to measure human performance on this task.}
We provide exactly the same set of operations to the user \Changed{}{in the GUI} as to the network, except for curve-based edits we provide $3$ control points instead of $8$ to make the interface more user-friendly. To introduce our software to the user, we show them a short \Changed{tutorial video}{video tutorial} before they start. $100$ images from $10$ users are collected, and their performance is given in Table~\ref{tab:benchmark_A}. \changed{}{Please see our supplemental document~\cite{hu2018b} for more details about data collection.} It can be seen from the user study that our method can generate results that are preferable to those produced by these users.

\section{Concluding Remarks}

Inspired by the retouching process of expert photographers, we proposed a general framework for automatic photo post-processing\changed{ that}{. It} utilizes \changed{}{three components:} {\em reinforcement learning} to reveal an understandable solution composed of common image manipulations, {\em generative adversarial networks} that allow training from unpaired image data, and {\em differentiable, resolution-independent filters} \changed{to make}{that make} network optimization possible over a variety of editing operators \changed{and for arbitrary image sizes}{on images of arbitrary resolution}. \changed{The effectiveness of this method}{Its effectiveness} was demonstrated through quantitative and qualitative comparisons. This framework is general enough to incorporate a broader set of operations, which we hope can make it even more versatile.

\begin{figure}
\includegraphics[width=0.15\textwidth]{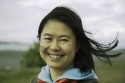}
\includegraphics[width=0.15\textwidth]{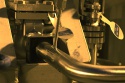}
\includegraphics[width=0.15\textwidth]{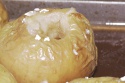}
\includegraphics[width=0.065\textwidth]{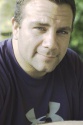}
\includegraphics[width=0.15\textwidth]{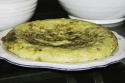}
\includegraphics[width=0.15\textwidth]{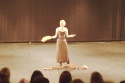}
  \caption{Example failure cases. Our method sometimes does not produce good tones for faces, as no special consideration is taken in our general framework of this particularly important aspect of photos. Also, our system may have limited ability to improve input photos that contain poor content, composition or lighting conditions.}
  \label{fig:failure}
\end{figure}

Certain low-level image filters, such as for pixel-level denoising, may be challenging to model as resolution-independent, differentiable filters, and thus may not fit into our framework. Without denoising, the image noise in shadows may become more pronounced after operations that boost brightness, as seen in Figure~\ref{fig:trajFhpB}. Denoising ideally should be applied to the input image prior to using our framework. Other failure cases are presented in Figure~\ref{fig:failure}.

\changed{}{Although we focused our discussions on global image operations, generalizing these operations to local ones would be a natural extension of this work. One straightforward way is to include parameters for spatially varying masks, which are used to blend the operation output with the input. Such masks can be analytically represented by functions that take as input the pixel location and intensity, parameterized by a few estimated parameters. Typical masks including ``gradient masks'' and ``luminance masks'' in {\em Photoshop} can be modeled in this way.}

For learning to retouch photos, we have only $2\times 10^3$ training images compared with the $1.4\times 10^7$ images in ImageNet for image classification. It would be meaningful in future work to 1) build larger datasets of RAW photos, and 2) transfer or reuse the knowledge distilled from ImageNet to the retouching problem.

In addition, it is possible to replace the actor-critic RL architecture and the Wasserstein GAN structure with other related alternatives. We find that much human labor and expertise is required to properly set the hyper-parameters to stabilize the training process. We believe that using more stable RL and GAN components will make this process easier and lead to even better results.

Finally, we hope that not only machines but also all interested people can understand the secrets of digital photography better, with the help of our ``Exposure'' system.



\if(0)
\newcommand{\GG}[2]{
  \includegraphics[width=.23\textwidth]{images/gallery3/#1/#2}
  \llap{
    \setlength{\fboxsep}{0pt}
    \color{white}\setlength{\fboxrule}{0.3pt}
    \fbox{\includegraphics[width=.06\textwidth]{images/gallery3/#1/#2}}
    \color{black}
  }
}

\newcommand{\GG}[2]{
  \def\stackalignment{l}
  \setlength{\fboxsep}{0pt}
  \color{white}\setlength{\fboxrule}{0.3pt}
  \topinset{\fbox{\includegraphics[width=.07\textwidth]{images/gallery3/#1/#2}}}{\includegraphics[width=.23\textwidth]{images/gallery3/#1/#2}}{0pt}{0pt}
  \color{black}
}
\fi

\clearpage
\newpage
\section*{\LARGE{Supplemental Document}}
\section{Filter Design Details}
\paragraph{Contrast, Saturation, and Black \& White Filters}
These filters are designed similarly, with an input parameter that sets the linear interpolation between the original image and the fully enhanced image, i.e.,

$$p_O=(1-p)
\cdot p_I+p \cdot \text{Enhanced}(p_I).$$

For {\textbf{Contrast}}:
$$\text{EnhancedLum}(p_I)=\frac{1}{2}(1-\cos(\pi \times (\text{Lum}(p_I)))),$$
$$\text{Enhanced}(p_I)=p_I \times \frac{\text{EnhancedLum}(p_I)}{\text{Lum}(p_I)},$$
where the luminance function $\text{Lum(p)}=0.27p_r + 0.67p_g + 0.06p_b.$

For {\textbf{Saturation}}:
$$\text{EnhancedS}(s, v)=s + (1 - s) \times (0.5 - |0.5 - v|) \times 0.8,$$
$$\text{Enhanced}(p_I)=\text{HSVtoRGB}(H(p_I), \text{EnhancedS}(S(p_I), V(p_I)). V(p_I)),$$
where $H$, $S$, and $V$ are HSV channels of a pixel.

For {\textbf{Black and White}}:
$$\text{Enhanced}(p_I)=\text{RGB}(\text{Lum}(p_I), \text{Lum}(p_I), \text{Lum}(p_I)).$$


\paragraph{Tone and Color Curves} We use a differentiable piecewise-linear mapping function to represent curves, as detailed in ~\cite{hu2018a}. For tone curves, the same curve is applied to the image, and the slope of each segment in the curve is in $[0.5, 2.0]$. For color, a separate curve is applied to each of the three color channels, with slopes in $[0.9, 1.1]$. The bounds on the curve slopes reflect the fact that human artists do not usually apply sharp color curves, but sometimes may use a strong tone curve.

\section{Experimental Details}

\paragraph{MIT-Adobe FiveK Dataset Partitions}

The MIT-Adobe FiveK dataset is randomly separated into three parts, which are listed in the data files {\bf FiveK\_train1.txt}, {\bf FiveK\_train2.txt} and {\bf FiveK\_test.txt}. For the test set, we select $100$ random images employed in the user study on AMT, as listed in file {\bf FiveK\_test\_AMT.txt}.

\paragraph{Histogram Intersection Details}
The quantities for histogram intersection are defined as follows:
\begin{itemize}
\item{Luminance} is defined as the mean pixel luminance (defined previously as $\textbf{Lum}.$)
\item{Contrast} is defined to be twice the variance of pixel luminance.
\item{Saturation} is defined as the mean pixel saturation (the ``S'' value in the HSL color space).
\end{itemize}

The results are separated into $32$ equal bins within the interval $[0, 1]$, i.e. $[0, 1/32), [1/32, 2/32), \ldots$

However, with only $1,000$ sample images, only about $31.25$ images will be placed in each bin on average, resulting in significant measurement noise. Therefore, we augment the data for histogram intersection by cropping $16$ patches in each image, and measure the histogram quantities on these $16,000$ image patches. Please refer to the accompanying code ({\bf histogram\_intersection.py}) for the detailed algorithm on measuring this error metric.

\begin{figure}[H]
\centering
    \includegraphics[width=.4\textwidth]{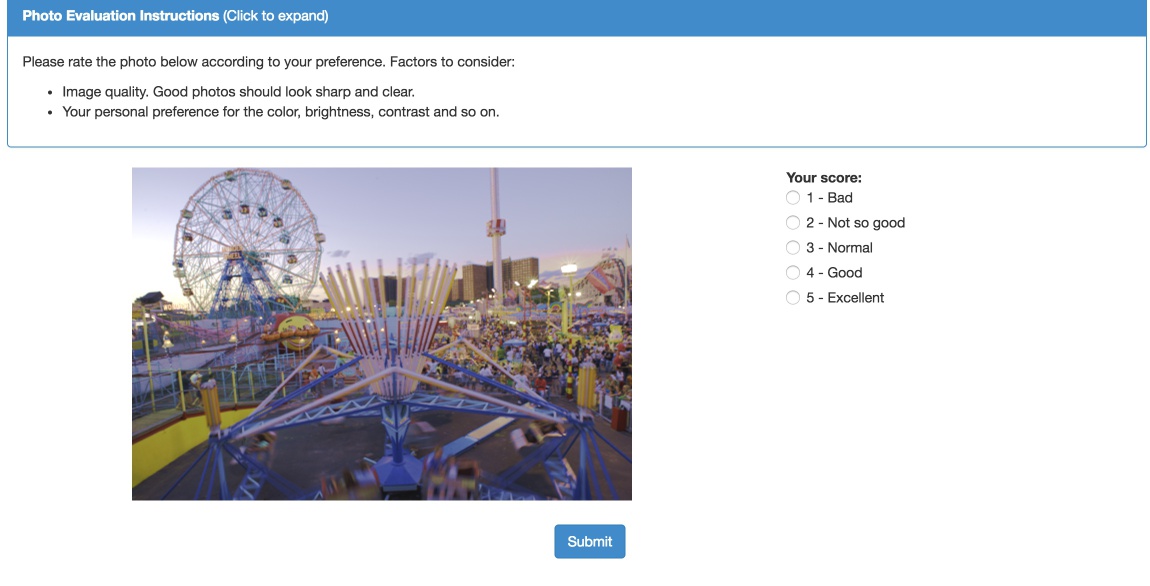}
    \includegraphics[width=.4\textwidth]{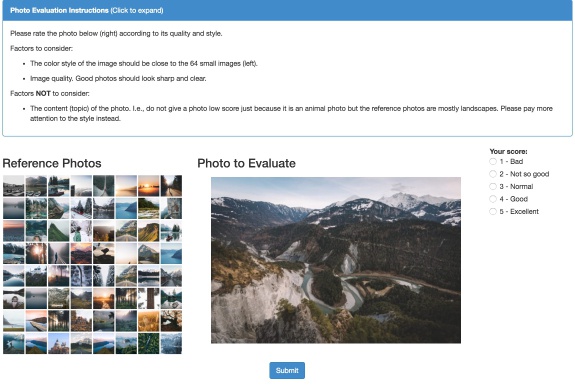}
    \caption{Our AMT UIs for user studies.}
    \label{fig:amt}
\end{figure}

\paragraph{Amazon Mechanic Turk} The AMT interfaces for evaluation are shown in Figure~\ref{fig:amt}.


\if(0)
\begin{figure}
\centering
    \includegraphics[width=1.0\linewidth]{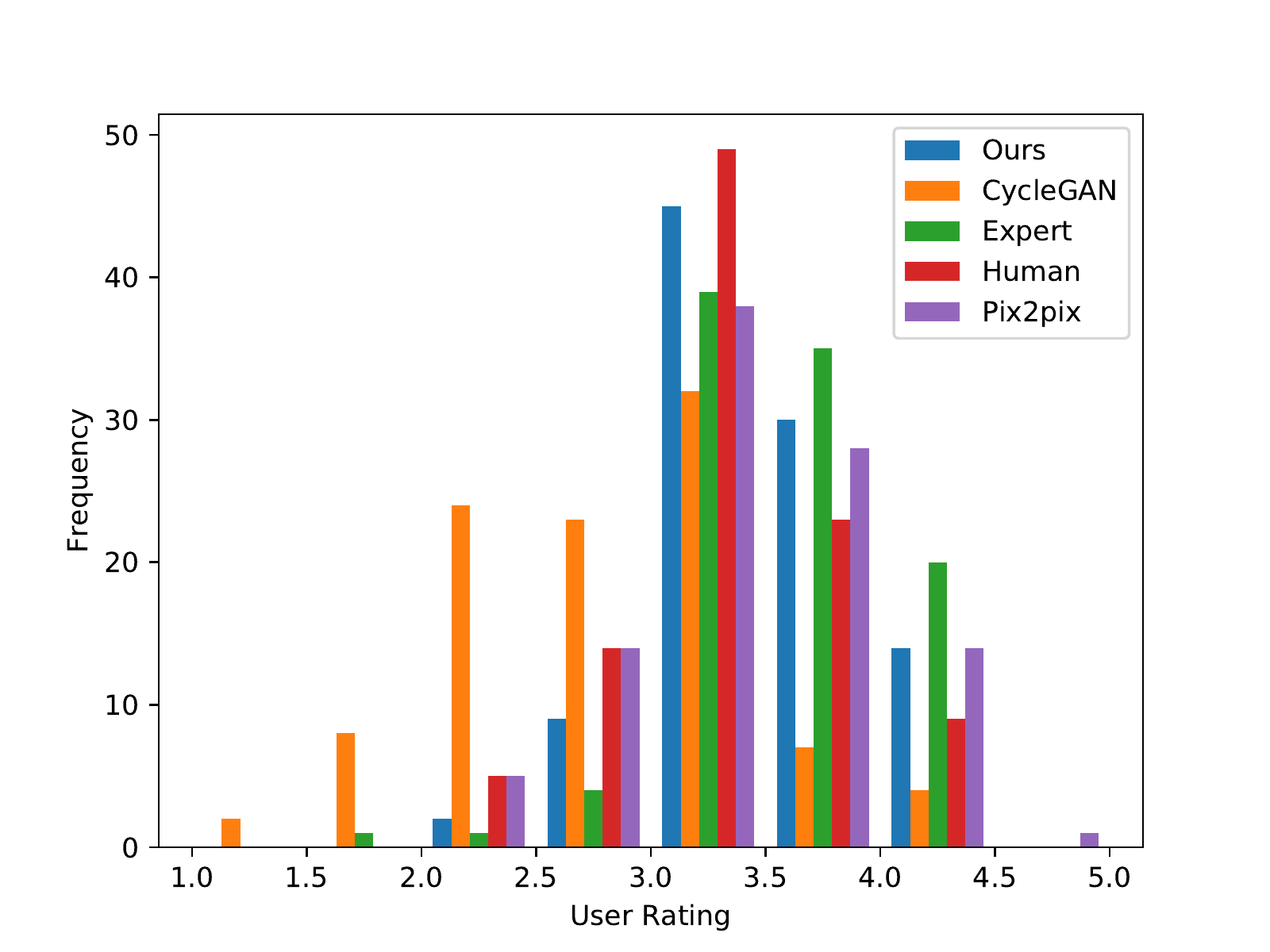}
    \caption{User rating distributions on general post-processing (using expert C in the MIT-Adobe FiveK dataset as training target dataset).}
    \label{fig:histC}
\end{figure}

\begin{figure}
\centering
    \includegraphics[width=1.0\linewidth]{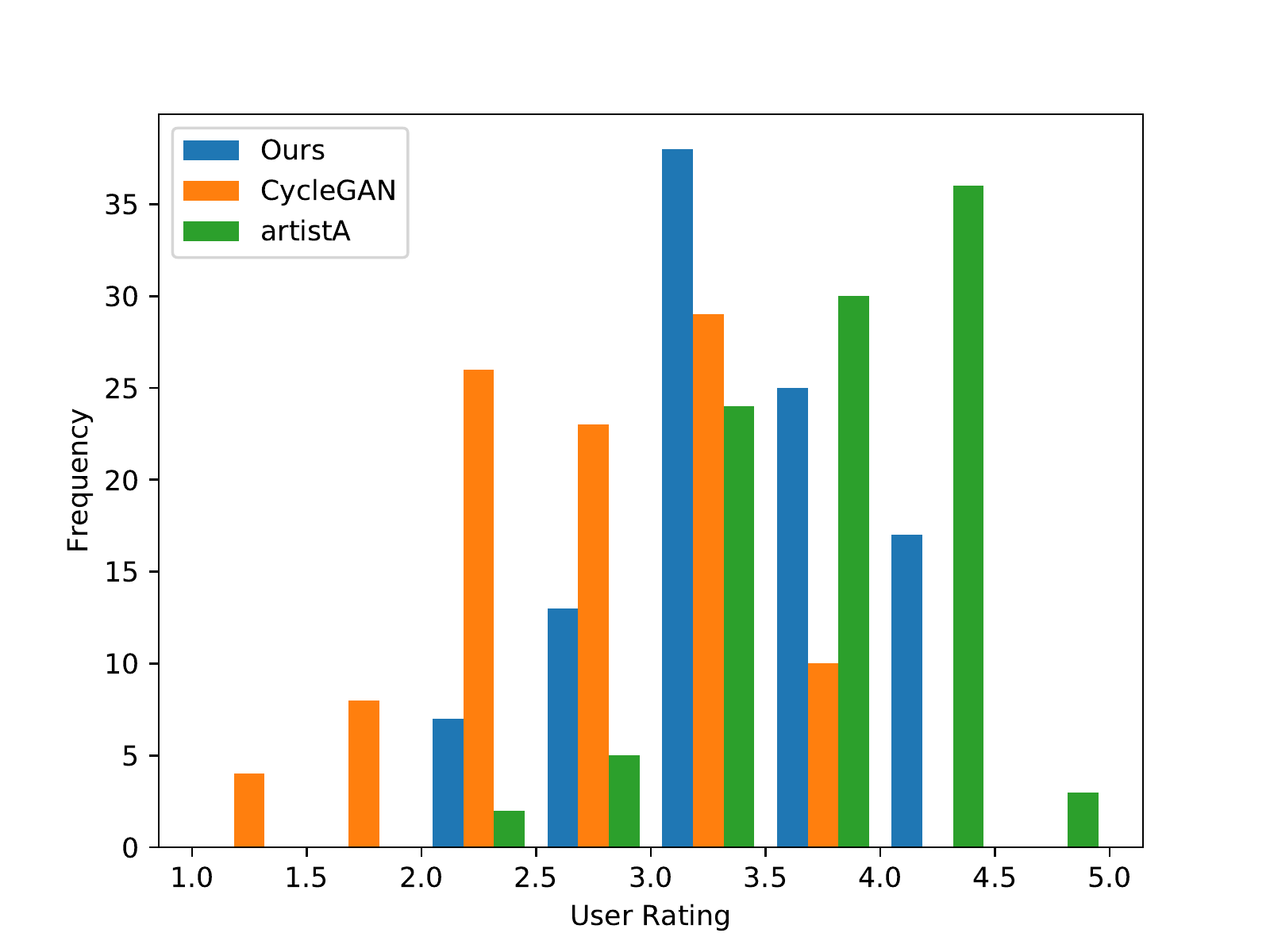}
    \caption{User rating distributions on style learning (using artist A in 500px as training target dataset).}
    \label{fig:histA}
\end{figure}

\begin{figure}
\centering
    \includegraphics[width=1.0\linewidth]{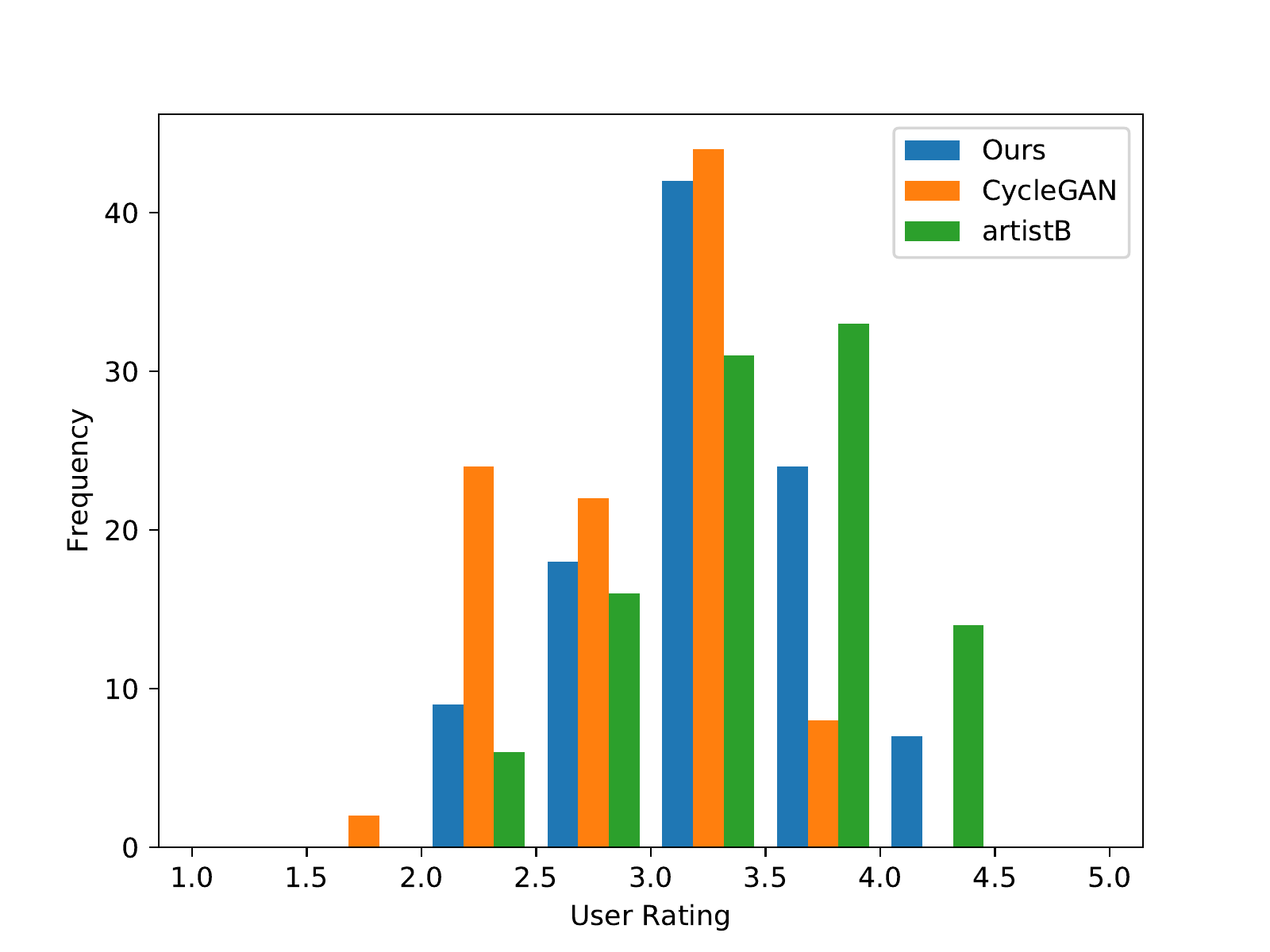}
    \caption{User rating distributions on style learning (using artist B in 500px as training target dataset).}
    \label{fig:histB}
\end{figure}
\fi

\begin{figure}
   \centering
    \includegraphics[width=0.45\textwidth]{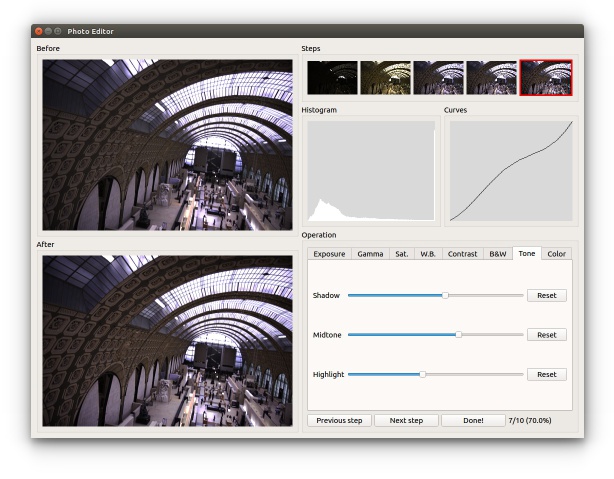}
  \caption{The graphical user interface for collecting retouching data from ordinary users. An intensity histogram of the current image and curves of color/tone curve operations are displayed.}
  \label{fig:ui}
\end{figure}

\paragraph{Human performance measurement}
\changed{}{A software with graphical user interface (Figure~\ref{fig:ui}) is developed to measure human performance on this task.}
We present the users a short video (with subtitles) demonstrating how our software should be used. The user studies take about $3$ minutes per image (roughly $30$ minutes for each user to retouch $10$ images). We do not enforce any time limit on the task. All $10$ users are highly educated and their ages range from $20$ to $30$.

\begin{figure*}[p]
\centering
 \includegraphics[width=.78\textwidth]{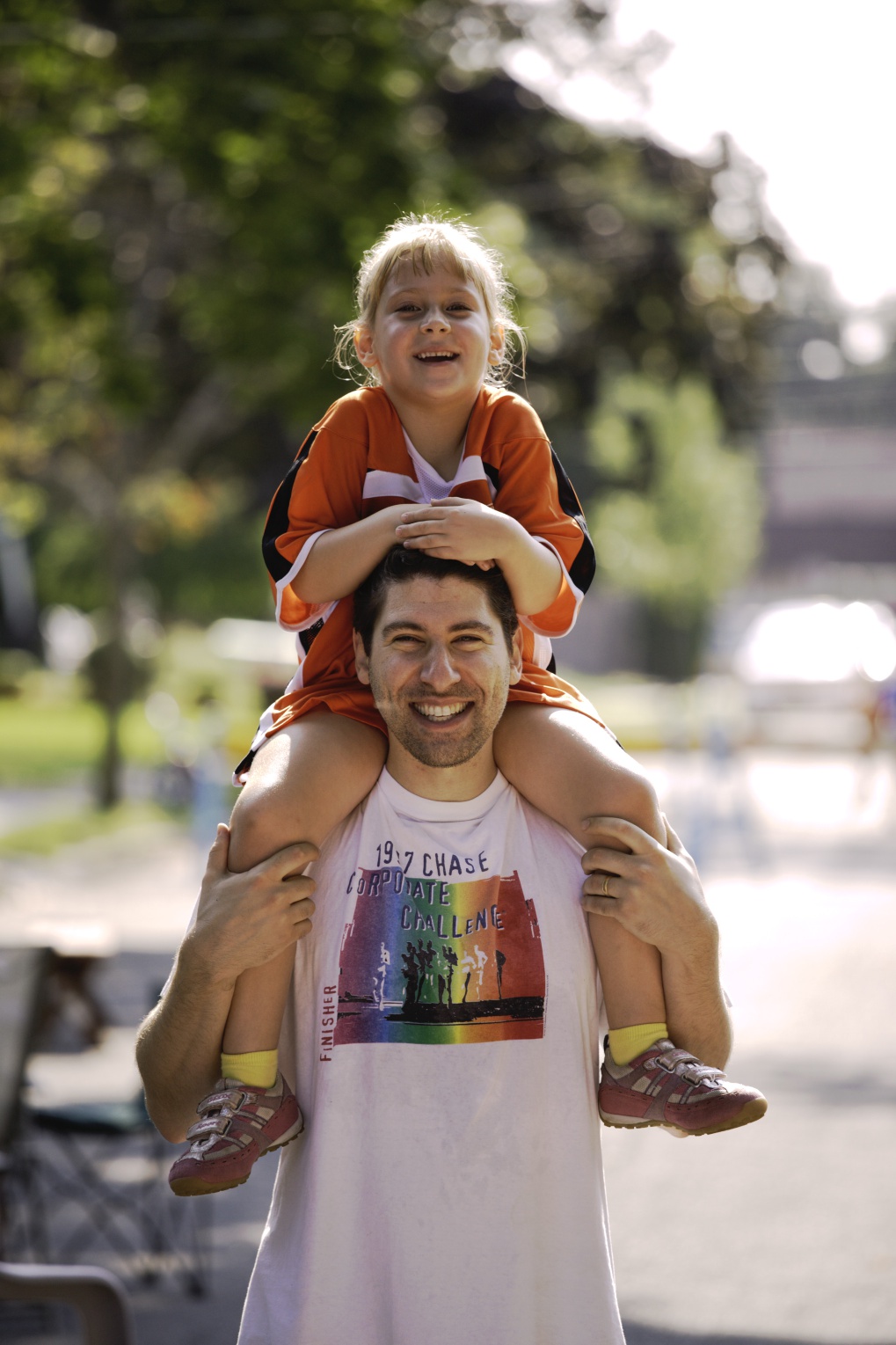}
    \caption{Our method can cleanly handle images of any resolution.}
    \label{fig:hd1}
\end{figure*}

\begin{figure*}[p]
\centering
 \includegraphics[width=.78\textwidth]{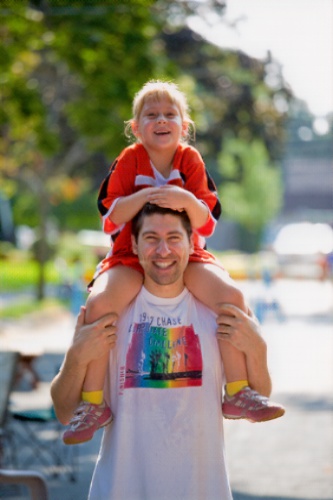}
    \caption{Pix2pix result.}
    \label{fig:hd2}
\end{figure*}

\begin{figure*}[p]
\centering
 \includegraphics[width=.78\textwidth]{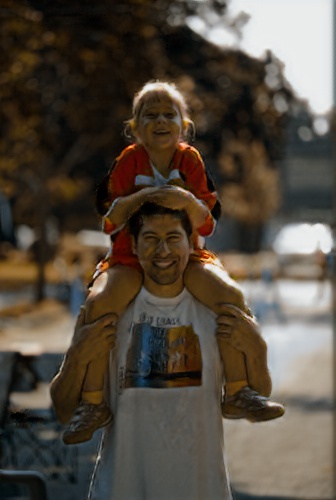}
    \caption{CycleGAN result.}
    \label{fig:hd3}
\end{figure*}

\paragraph{Scalability in Resolution}
The ability to process high-resolution images is critical in professional photography. In Figure~\ref{fig:hd1}, ~\ref{fig:hd2} and ~\ref{fig:hd3}, we show high-resolution results from our method, Pix2pix, and CycleGAN. It is clear that our method produces images with the highest quality on high-resolution images.

\clearpage
\newpage

\bibliographystyle{ACM-Reference-Format}

\bibliography{main}

\end{document}